\documentclass[aps,nofootinbib,superscriptaddress, showpacs,preprintnumbers,  nofootinbibt,twocolumn]{revtex4-2}
\usepackage{epsfig}
\usepackage{multirow}
\usepackage{eurosym}
\usepackage{dcolumn}
\usepackage{bm}
\usepackage{enumerate}
\usepackage{float}
\usepackage{epstopdf}
\usepackage{amsmath}
\usepackage{bm}
\usepackage{amsfonts}
\usepackage{amssymb}
\usepackage{graphicx}
\usepackage{alphalph,mathtools}
\usepackage{etoolbox}
\usepackage{color}
\usepackage{booktabs}
\usepackage{footnote}
\usepackage{makecell,tabularx}
\usepackage{hyperref}
\hypersetup{colorlinks,citecolor=blue}
\hypersetup{colorlinks=true,linkcolor=magenta,filecolor=magenta,urlcolor=blue}

\def\be{\begin{equation}}
    \def\ee{\end{equation}}
\def\bea{\begin{eqnarray}}
    \def\eea{\end{eqnarray}}

\begin{document}

\title{Constraints on the Parameterized Deceleration Parameter in FLRW Universe}
\author{Himanshu Chaudhary}
\email{himanshuch1729@gmail.com} 
\affiliation{Department of Applied Mathematics, Delhi Technological University, Delhi-110042, India,}
\affiliation{Pacif Institute of Cosmology and Selfology (PICS) Sagara, Sambalpur 768224, Odisha, India}
\affiliation{Department of Mathematics, Shyamlal College, University of Delhi, Delhi-110032, India,}
\author{Amine Bouali}
\email{a1.bouali@ump.ac.ma} 
\affiliation{Laboratory of Physics of Matter and Radiation, Mohammed I University, BP 717, Oujda, Morocco,}
\author{Ujjal Debnath}
\email{ujjaldebnath@gmail.com} \affiliation{Department of
Mathematics, Indian Institute of Engineering Science and
Technology, Shibpur, Howrah-711 103, India,}
\author{Tanusree Roy}
\email{tanusreeroy1995@gmail.com} \affiliation{Department of
Mathematics, Indian Institute of Engineering Science and
Technology, Shibpur, Howrah-711 103, India,}
\author{G.Mustafa}
\email{gmustafa3828@gmail.com} \affiliation{Department of Physics,
Zhejiang Normal University, Jinhua 321004, People’s Republic of
China,}
\affiliation{New Uzbekistan University, Mustaqillik ave. 54, 100007 Tashkent, Uzbekistan,}

\begin{abstract}
Confirmation of accelerated expansion of the universe probed the
concept of dark energy theory, and since then, numerous models
have been introduced to explain its origin and nature. The present
work is based on reconstructing dark energy by parametrization of
the deceleration parameter in the FLRW universe filled with
radiation, dark matter and dark energy. We have chosen some
well-motivated parametrized models 1-3 in an attempt to
investigate the energy density in terms of deceleration parameters
by estimating the cosmological parameters with the help of
different observational datasets. Also, we have introduced a new
model 4 for the parametrization of the deceleration parameter.
Then we analyzed the cosmography parameters using the best-fit
values of the parameters. Using the information criteria, we have
examined the viability of the models.
\end{abstract}

\maketitle
\tableofcontents


\section{Introduction}\label{Introduction}
The concept of cosmic acceleration was probably one of the most
promising discoveries in the modern cosmology paradigm. Recently,
two independent research works involving distant supernovae
suggested that in the present epoch, the universe is undergoing an
accelerated expansion \cite{1,2}. This phenomenon has been
favorably explained later by the existence of an energy component
with massive negative pressure comprising nearly 70\% of the
universe. This is known as  ``dark energy" (DE). The nature of
this is still unidentified. Synchronizing with the observed data,
many DE models have been proposed so far. Among them, the
$\Lambda$CDM model is widely accepted as supposedly it `best
accommodates' the observations but also it comes with some
disadvantages like a fine-tuning problem, coincidence problems, and
so on \cite{3,4,5}. To overcome these drawbacks, alternative DE
models have been explored like a quite favorable phantom,
$k$-essence, Chaplygin gas, etc, \cite{6} for possible
explanations of the origin and nature of the dark energy. However,
prior to the accelerated phase, the universe had gone through a
decelerated phase in the early epoch where the effects of dark
energy were absent or subdominant, some recent late-time cosmic acceleration is discussed in \cite{koussour2022anisotropic,koussour2022lambda,koussour2023constant,chaudhary2023cosmological}. It is believed that density
perturbations occurred in this epoch which played a key role in
the structural formation of the universe. So, to cover the entire
evolution, we would have to employ a cosmological model which
would simultaneously describe the accelerated and decelerated
phases. Earlier, the definition of cosmology was addressed by
Sandage\cite{7} as a search for two simple but fundamental
cosmographic parameters: Hubble parameter ($H_0$), which
determined the expansion rate and a small correction $q_0$ due to
gravity, known as deceleration parameter (DP), responsible for
slowing down the expansion. Though the inclusion of `dark energy
has completely changed the scenario, but still, any practical
aspect of cosmological evolution is tightly bound to DP. It is
defined as $q=-\frac{a\ddot{a}}{\dot{a}^2}$ where $a(t)$ is the
scale factor of the universe. $q>0$ ($\ddot{a}<0$) indicates a
decelerating universe and vice versa. While HP describes the
linear part of the time dependence of the scale factor, the
non-linear correction term $q_0$ opens up possibilities like the
presence of local instabilities or the existence of chaotic
regimes \cite{8}. Moreover, the dynamics of observable galaxy
number variation can be determined through DP. Like DE cosmological models, DE has been subjected to numerous modifications to be better fitted with observational data.
\\

Parametrization of DP as a function of scale factor $a$ or
redshift $z$ can be accounted as a suitable approach to it.
Limitations to such parametric assumptions are: (i) Most of the
parametrizations diverge in the distant future and some of them
are only valid at low redshift limit ($z<<1$) \cite{9,10}; (ii) Prior parametric assumptions may be in conflict with the true nature of the dark energy. (iii) In non-parametric models, evolution can be directly deduced from observational data avoiding parametric assumptions \cite{11,12,13,14,15}. However, it can help to improve the efficiency of future cosmological surveys. So in pursuit of
understanding the transition from decelerated to accelerated
phase, the parametrization approach can be proved fruitful.
Recently Capozziello reconstructed a divergence-free form of DP
starting with Pade polynomials and analyzed the corresponding
observational data \cite{16}. A logarithmic parametrization of DP
was proposed in ref. \cite{17}, and the constraints were obtained
by using type Ia supernova, BAO, and CMB data sets. Motivated by
these ideas, we have adopted some well-motivated parametrizations
of DP to reconstruct dark energy and, consequently, Hubble
parameter $H(z)$ in terms of redshift $z$. Mainly,
well-established parametrized models have been introduced for dark
energy equation of state and constrain the model parameters by
observational data analysis. In the study of the generalized
holographic dark energy model, some well-known parametrization
type models have been considered \cite{18,19}. Till now, some
authors have assumed some possible forms of parametrization of
deceleration parameter \cite{20,21,22,23,24,25,bouali2023data,boualicosmological,bouali2023model}. The main advantage behind introducing aparameterized deceleration parameter is to provide a framework that is independent of specific gravitational theories. By parameterizing the deceleration parameter, we can explore the behavior of cosmic expansion without being tied to a particular model. This flexibility allows us to investigate a wide range of cosmological scenarios and potentially uncover new physics beyond our current understanding. Furthermore, in the well-established cosmological models, such as the $\Lambda$CDM model, the focus has primarily been on parameterizing the equation of state of dark energy. However, in the referenced papers \cite{26,27}, the authors have recognized the importance of considering the deceleration parameter as a viable alternative. This motivates us to extend the analysis to include the parametrization of the deceleration parameter for these well-established models and constrains the model parameters by employing MCMC data analysis, which is a powerful statistical technique widely used in cosmology. This approach enables us to explore the parameter space efficiently and extract robust constraints by comparing the theoretical predictions of the models with observational data. Our utilization of MCMC analysis adds a robust statistical framework to our study, enhancing the reliability of our results.  A key contribution of our work is the comprehensive comparison of different models with the standard $\Lambda$CDM model. The $\Lambda$CDM model has been highly successful in explaining various cosmological observations, and serves as the benchmark against which we assess the viability of alternative models. By quantitatively evaluating the goodness-of-fit and model selection criteria, we can determine which models provide a better description of the observational data, thereby highlighting their relative viability.
\\

The main focus of the work is to constrain the model parameters
using recently released data. Here, in particular, we have chosen
to use the updated astronomical datasets: the measurements of
Hubble parameter from the differential evolution of cosmic
chronometers (CC); SNIa datasets from Type Ia Supernovae
sample comprising 1048 measurements; 17 measurements of baryons
acoustic oscillation (BAO) data. New constraints on DP have been
provided by jointly analyzing the above datasets and implementing
Monte Carlo Markov Chain (MCMC) method. The era of modern
cosmology promotes the study of kinematic quantities, vastly known
as "Cosmography" or "Cosmo-kinetics". The very idea of it is
observationally driven and completely independent of any prior
assumption of the gravity theory or elected cosmological model.
Cosmography presents itself with a compelling advantage as it
simply follows the symmetry principles and direct observation-
without involving Einstein's equations (Friedmann equations).
Consequently, we can steadfastly avoid some arguable speculations
regarding 'dark energy', 'dark matter', and others. While pure
cosmography does not envision the scale factor $a(t)$ itself but
the history of its evolution can be inferred to some extent.
Dunajski and Gibbons \cite{33} have studied the constraints on the
cosmographic parameters like the deceleration, jerk, and snap
parameters for different dark energy models. The parameterization
of these quantities is discussed in \cite{34,35,36,37}. Shafieloo,
Kim and Linder \cite{22} have discussed the non-parametric reconstruction of these quantities.\\

The organization of the paper is as follows: In section \ref{sec1}, we
consider the basic equations of the FLRW model. The Hubble
parameter is written in terms of the deceleration parameter. We consider parametrized deceleration parameter models
like models 1, 2, 3, and 4. In section \ref{sec2}, Section \ref{sec3} deals with the data descriptions like cosmic chronometric datasets, SNIa datasets, and BAO datasets with MCMC results. In
section \ref{sec5} we fit the models with $H(z)$ and SNIa datasets. In section \ref{sec6}, we discuss the cosmography parameters. 
In section \ref{sec7}, we analyze the detailed description of the model parameters. In section \ref{sec8}, we present the information criteria for our models. Finally, the results are presented in section \ref{sec9}.

\section{Basic Equations of FLRW Model}\label{sec1}
We have considered a spatially flat, homogeneous, isotropic FLRW
the universe with line element

\begin{eqnarray}
ds^2=-dt^2 +a^2(t)\left[dr^2+r^2\left(d\theta^2 + sin^2 \theta
d\phi^2 \right)\right]
\end{eqnarray}

$a(t)$ being the scale factor.

The energy-momentum tensor of the fluid reads as

\begin{eqnarray}
T_{\mu\nu}=(\rho+p)u_\mu u_\nu+p g_{\mu\nu}
\end{eqnarray}

where $\rho$ and $p$ are the energy density and pressure density
of the fluid
respectively. The fluid 4-velocity $u^\mu=\frac{dx^\mu}{ds}$ satisfies the relation $u^\mu u_\mu=-1$.\\

For the FLRW Universe, the Friedmann equations in Einstein's gravity
are given by
\begin{equation}\label{F1}
H^2=\frac{8\pi G}{3}~\rho
\end{equation}
and
\begin{equation}\label{F2}
\dot{H}=-4\pi G(\rho+p)
\end{equation}
where, $H=\dot{a}/a$ is the Hubble parameter and overhead dot
denotes derivative with cosmic time $t$. Considering that the
the universe is filled with fluid matter of total energy density
$\rho$ and total pressure $p$, it obeys the energy conservation
equation
\begin{equation}\label{Cons}
\dot{\rho}+3H(\rho+p)=0
\end{equation}

We start with the prediction that the universe is composed of
matter content comprising radiation, dark matter (DM) and dark
energy (DE). So, $\rho$ and $p$ consist of densities and pressures
of radiation, DM and DE. So $\rho=\rho_{r}+\rho_{m}+\rho_{d}$ and
$p=p_{r}+p_{m}+p_{d}$. Now assume that radiation, DM and DE
follows the conservation equation separately so that

\begin{equation}\label{rad}
\dot{\rho}_{r}+3H(\rho_{r}+p_{r})=0,
\end{equation}

\begin{equation}\label{DM}
\dot{\rho}_{m}+3H(\rho_{m}+p_{m})=0
\end{equation}

and

\begin{equation}\label{DE}
\dot{\rho}_{d}+3H(\rho_{d}+p_{d})=0
\end{equation}

For radiation, $p_{r}=\frac{1}{3}\rho_{r}$, so from equation
(\ref{rad}) we obtain $\rho_{r}=\rho_{r0}a^{-4}$. Since the DM
follows negligible pressure (i.e., $p_{m}=0$), so from equation
(\ref{DM}) we obtain $\rho_{m}=\rho_{m0}a^{-3}$.

Let us consider the deceleration parameter
\begin{eqnarray}\label{q0}
q=-1-\frac{\dot{H}}{H^2}
\end{eqnarray}
The corresponding deceleration parameter for DE is given by

\begin{eqnarray}
q_{d}=-1-\frac{\dot{H_{d}}}{H_{d}^2}
\end{eqnarray}
where $H_{d}$ is the Hubble expansion rate of the dark energy
term. So from equations (\ref{F1}) and (\ref{F2}), we can write
\begin{equation}\label{F11}
H_{d}^2=\frac{8\pi G}{3}~\rho_{d}
\end{equation}
and
\begin{equation}\label{F22}
\dot{H}_{d}=-4\pi G(\rho_{d}+p_{d})
\end{equation}

Using the field equations \eqref{F11} and \eqref{F22} and the
energy-conservation equation \eqref{DE}, the fluid energy density
becomes
\begin{equation} \label{rho}
\rho_{d}=\rho_{d0}~ e^{\int \frac{2(1+q_{d})}{1+z}dz}
\end{equation}
where $\rho_{d0}$ represents the present value of the density
parameter and $z$ is the redshift parameter described as
$1+z=\frac{1}{a}$ (presently, $a_{0}=1$). Defining
$\Omega_{r0}=\frac{8\pi G\rho_{r0}}{3H_{0}^{2}}$,
$\Omega_{m0}=\frac{8\pi G\rho_{m0}}{3H_{0}^{2}}$ and
$\Omega_{d0}=\frac{8\pi G\rho_{d0}}{3H_{0}^{2}}$, from equation
(\ref{F1}), we obtain the Hubble parameter as

\begin{equation}\label{H}
H^{2}(z) =H_{0}^{2}\left[\Omega_{r0}(1+z)^{4}+\Omega_{m0}(1+z)^{3}
+ \Omega_{d0}~ e^{\int \frac{2(1+q_{d})}{1+z}dz}\right]
\end{equation}

where $\Omega_{d0}=1-\Omega_{r0}-\Omega_{m0}$.\\

To find the deceleration parameter $q$ using the expression for the Hubble parameter $H(z)$, we first need to calculate the derivative of $H(z)$ with respect to $z$. Then we substitute this derivative into the formula for $q$. Differentiating both sides of the equation with respect to $z$, we have:

\begin{equation}
\begin{aligned}
2H(z)\frac{dH(z)}{dz} &= H_{0}^{2}\left[4\Omega_{r0}(1+z)^{3} + 3\Omega_{m0}(1+z)^{2}\right. \\
&\quad+ \left.\frac{2(1+q_{d})\Omega_{d0}}{(1+z)}e^{\int \frac{2(1+q_{d})}{1+z}dz}\right]
\end{aligned}
\end{equation}

Now, we can substitute this derivative of $H(z)$ into the expression for $q$:

\begin{equation}
q = -1 - \frac{\dot{H}}{H^2} = -1 - \frac{2H(z)\frac{dH(z)}{dz}}{H(z)^2}
\end{equation}

Substituting the expression for $\frac{dH(z)}{dz}$ derived earlier, we get:

\begin{equation}
\begin{aligned}
q = -1 - \frac{2}{H(z)} \bigg[ & 4\Omega_{r0}(1+z)^{3} + 3\Omega_{m0}(1+z)^{2} \\
& + \frac{2(1+q_{d})\Omega_{d0}}{(1+z)}e^{\int \frac{2(1+q_{d})}{1+z}dz} \bigg]
\end{aligned}
\end{equation}

Now, one could evaluate $H(z)$ using the given values of $\Omega_{r0}$, $\Omega_{m0}$, $\Omega_{d0}$, and $q_d$ at the specific redshift $z$ you are interested in, and substitute it into the equation to calculate the corresponding deceleration parameter $q$.
\\

For $\Lambda$CDM model, by adding the cosmological constant $\Lambda$ terms in equations (\ref{F1}) and (\ref{F2}) and by assuming the density parameter $\Omega_{\Lambda 0}=\frac{\Lambda}{3H_0^2}$, the deceleration parameter $q$ from equation (\ref{q0}) can be written as
\begin{equation}
q(z)=\frac{\left[2\Omega_{r0}(1+z)^{4}+\Omega_{m0}(1+z)^{3}-2\Omega_{\Lambda 0}\right] }{ 2\left[\Omega_{r0}(1+z)^{4}+\Omega_{m0}(1+z)^{3}+\Omega_{\Lambda 0}\right]}    
\end{equation}

where $\Omega_{\Lambda 0}=1-\Omega_{r0}-\Omega_{m0}$.
The present value of the deceleration parameter is found
by inserting $z = 0$, which gives $q(z=0)=\frac{\left[2\Omega_{r0}+\Omega_{m0}-2\Omega_{\Lambda 0}\right] }{ 2\left[\Omega_{r0}+\Omega_{m0}+\Omega_{\Lambda 0}\right]}~.$\\

\section{Parameterized deceleration parameter}\label{sec2}

Most simplest parametrization of $q$ which contains two parameters
can be taken as

\begin{equation}
q(z)=q_0+q_1 \mathcal{X}(z)
\end{equation}

where $q_0$ and $q_1$ are constants and $\mathcal{X}(z)$ is a
function of redshift $z$. In search of satisfactory solutions to
the cosmological puzzles, many forms of $\mathcal{X}(z)$ has been
suggested. As mentioned earlier, most of them were inadequate in
explaining future evolution scenarios. So the persuasion of an
ideal divergence-free parametrization of DP is still relevant. The
well-known parametrized equation of state parameter models has
been introduced by several authors, and the corresponding
analogous of these models for parametrized deceleration parameters
have been introduced in \cite{26,27}. Here, we have adopted the
analogous of some well-known parametrized models for the
deceleration parameter, which contains two unknown parameters and
calculated the corresponding Hubble parameter in terms of redshift
$z$.

\subsection{Model 1 (Wetterich type)}

The Wetterich model for the parametrized equation of state parameter
has been studied in \cite{38,39}. The analogous Wetterich type
parametrization of the deceleration parameter has been introduced in
\cite{26,27} and is given by

\begin{equation}\label{Wetterich}
q_{d}(z)=\frac{q_0}{1+q_1 1og(1+z)}
\end{equation}

where $q_{0}$ and $q_{1}$ are constants. Then the energy density
will be given by

\begin{equation}\label{rho Wetterich}
\rho_d=\rho_{d0}~(1+z)^{2}~\left\{1+q_1~
log(1+z)\right\}^{\frac{2q_{0}}{q_{1}}}
\end{equation}

From equation (\ref{H}), we obtain

\begin{equation}
\begin{aligned}
H^2(z) & =H_0^2\left[\Omega_{r 0}(1+z)^4+\Omega_{m 0}(1+z)^3\right. \\
& \left.+\left(1-\Omega_{r 0}-\Omega_{m
0}\right)(1+z)^2\left\{1+q_1 \log (1+z)\right\}^{\frac{2 q_0}{q
1}}\right]
\end{aligned}
\end{equation}

\subsection{Model 2 (Barboza-Alcaniz type)}

The Barboza-Alcaniz model for parametrized equation of state
parameter has been studied in \cite{44}. The analogous
Barboza-Alcaniz type parametrization of deceleration parameter has
been introduced in \cite{26,27} and is given by

\begin{equation}\label{Bar}
q_{d}(z)=q_0+q_1\frac{z(1+z)}{1+z^2}
\end{equation}

where $q_{0}$ and $q_{1}$ are constants. Then the energy density
will be

\begin{equation}\label{rho Bar}
\rho_d=\rho_{d0}~(1+z)^{2(1+q_0)}~\left(1+z^2\right)^{q_{1}}
\end{equation}

From equation (\ref{H}), we obtain

\begin{equation}
\begin{aligned}
H^2(z) & =H_0^2\left[\Omega_{r 0}(1+z)^4+\Omega_{m 0}(1+z)^3\right. \\
& \left.+\left(1-\Omega_{r 0}-\Omega_{m
0}\right)(1+z)^{2(1+q_0)}~\left(1+z^2\right)^{q_{1}}\right]
\end{aligned}
\end{equation}

\subsection{Model 3 (CPL type)}

The famous Chevallier-Polarski-Linder (CPL) model for parametrized
equation of state parameter has been studied in \cite{34,35}. The
analogous CPL type parametrization of deceleration parameter has
been introduced in \cite{26,27} and is given by
\begin{equation}\label{CPL}
    q_{d}(z)=q_0+q_1 \frac{z}{1+z}
\end{equation}

where $q_{0}$ and $q_{1}$ are constants. Subsequently, the energy
density \eqref{rho} becomes

\begin{equation}\label{rho CPL}
    \rho_d=\rho_{d0}~(1+z)^{2(1+q_0+q_1)}~e^{\frac{2q_1}{1+z}}
\end{equation}

From equation (\ref{H}), we obtain

\begin{equation}
\begin{aligned}
H^2(z) & =H_0^2\left[\Omega_{r 0}(1+z)^4+\Omega_{m 0}(1+z)^3\right. \\
& \left.+\left(1-\Omega_{r 0}-\Omega_{m
0}\right)(1+z)^{2(1+q_0+q_1)}~e^{\frac{2q_1}{1+z}}\right]
\end{aligned}
\end{equation}

\subsection{Model 4}
Here we propose a new parametrized model for deceleration
parameter and is given as in the form:

\begin{equation}\label{New}
    q_{d}(z)=q_0+q_1 \frac{1+z}{2+z}
\end{equation}

where $q_{0}$ and $q_{1}$ are constants. Subsequently, the energy
density \eqref{rho} becomes

\begin{equation}\label{rho New}
    \rho_d=\rho_{d0}~(1+z)^{2(1+q_0)}(2+z)^{2q_{1}}
\end{equation}

From equation (\ref{H}), we obtain

\begin{equation}
\begin{aligned}
H^2(z) & =H_0^2\left[\Omega_{r 0}(1+z)^4+\Omega_{m 0}(1+z)^3\right. \\
& \left.+\left(1-\Omega_{r 0}-\Omega_{m
0}\right)(1+z)^{2(1+q_0)}(2+z)^{2q_{1}}\right]
\end{aligned}
\end{equation}

\section{Data Analysis}\label{sec3}
In this section, we will constrain our model parameters by using
three types of dataset. The CC datasets consist 31 measurements,
The SNIa dataset consists 1048 measurements and 17 measurements of 
BAO to obtain the best-fit value of our model parameters. We have implemented the Markov Chain Monte Carlo (MCMC) \cite{28emcee} and implemented with the open-source package Polychord \cite{29polychord} and GetDist \cite{30getdist}. The total $\chi^2$ function of the combination CC + BAO + SNIa and define as

$$
\chi^{2}=\chi_{CC}^{2} + \chi_{SNIa}^{2} + \chi_{B A O}^{2}.
$$

\subsection{Data description}\label{Observation}

\subsubsection{Cosmic Chronometric (CC) datasets}
We consider the compilation of 31 measurements of CC lying between
the redshift range $0.07 \leq z \leq 1.965$. The underlying
principle for these measurements was proposed in \cite{31}, by
relating the Hubble parameter with redshift $z$, and cosmic time
$t$

$$
H(z)=-\frac{1}{1+z} \frac{\mathrm{d} z}{\mathrm{~d} t}
$$

The $\chi^2$ function for these measurements, denoted by

$\chi_{\mathrm{CC}}^{2}$, is

\begin{equation}
\chi_{\mathrm{CC}}^{2}=\sum_{i=1}^{31} \frac{\left[H^{t
h}\left(z_i\right)-H^{o b
s}\left(z_i\right)\right]^2}{\sigma_{H^{\mathrm{obs}}\left(z_{i}\right)}^{2}},
\end{equation}

where $H^{\mathrm{th}}\left(z_{i}, k,\alpha, h\right)$ represent
the theoretical value obtained from our cosmological model,
$H^{\text {obs }}\left(z_{i}\right)$ and represent the observed
value of hubble parameter with standard deviation
$\sigma_{H^{\mathrm{obs}}\left(z_{i}\right)}^{2}$. (to see more
and rundown all measurements see \cite{32bouali2019cosmological})

\subsubsection{type Ia supernova (SNIa) datasets}

The 1048 measurements of type Ia supernovae from five different sub-samples SNLS, SDSS, PSI, low-$z$, and HST in the redshift range of $0.01<z<2.3$ \cite{zhai2019robust}. The $\chi^{2}$ function of the SNIa data is given as

\begin{equation}
\chi_{\mathrm{SNIa}}^2=\Delta \mu C_{\mathrm{Pan}}^{-1} \Delta
\mu^T,
\end{equation}

where $\Delta \mu = \mu_i^{\text {obs }}-\mu^{\text {th }}$. Where
$\left(\mu_i^{\text {obs }}\right)$ represented as observed
distance modulus and evaluated as

\begin{equation}
\mu_i^{\text {obs }}=\mu_{B, i}+\mathcal{M},
\end{equation}

$\mu_{B, i}$ represents the observed peak magnitude at maximum in
the rest frame of the $B$ band for redshift $z_i$, while
$\mathcal{M}$ represents nuisance parameter. The theoretical
distance modulus was evaluated as

\begin{equation}
\mu^{\text {th }}=5 \log _{10} D_L+\mathcal{M},
\end{equation}

where

\begin{equation}
D_L=\left(1+z_{\mathrm{hel}}\right) \int_0^{z_{\mathrm{cmb}}}
\frac{H_0 d z}{H(z)},
\end{equation}

with $z_{\text {hel }}$ is heliocentric and $z_{\mathrm{cmb}}$ is
CMB rest frame redshifts. The covariance matrix is measured as
$C_{\text {Pan }}=C_{\text {sys }}+D_{\text {stat }}$. Where
$C_{\text {sys }}$ is the systematic covariance matrix and $D_{\text
{stat }}$ stands for diagonal of the covariance matrix of the
statistical uncertainty and is calculated as

\begin{equation}
D_{\text {stat }, i i}=\sigma_{\mu_{B, i}}^2 .
\end{equation}

The description and the systematic covariance matrix together with
$\mu_{B, i}, \sigma_{\mu_{B, i}}^2, z_{\mathrm{cmb}}$, and
$z_{\text {hel }}$ for the $i$ th SnIa are mentioned in
\cite{scolnic2018complete}.

\subsubsection{Baryon Acoustic Oscillations} We picked 17 BAO \cite{benisty2021testing}
measurements from the greatest collection of BAO dataset of (333)
measurements since adopting the entire catalog of BAO might lead
in a very significant error owing to data correlations, therefore
we opted for a small dataset to minimize inaccuracies. Transverse BAO
studies contribute measurements of $D_H(z) / r_d=c / H(z)
r_d$ with comoving angular diameter
distance.\cite{hogg2020constraints} \cite{martinelli2020euclid}.

\begin{equation}
D_M=\frac{c}{H_0} S_k\left(\int_0^z \frac{d
z^{\prime}}{E\left(z^{\prime}\right)}\right) \text {, }
\end{equation}

where

\begin{equation}
S_k(x)= \begin{cases}\frac{1}{\sqrt{\Omega_k}} \sinh
\left(\sqrt{\Omega_k} x\right) & \text { if } \quad \Omega_k>0 \\
x & \text { if } \quad \Omega_k=0 \\ \frac{1}{\sqrt{-\Omega_k}}
\sin \left(\sqrt{-\Omega_k} x\right) & \text { if } \quad
\Omega_k<0 .\end{cases}
\end{equation}

We also consider the angular diameter distance $D_A=$ $D_M /(1+z)$
and the $D_V(z)/r_d$. which is combination of the BAO peak
coordinates and $r_d$ is the sound horizon at the drag epoch.
Finally we can obtain "line-of-sight" (or "radial") observations
directly the Hubble parameter

\begin{equation}
D_V(z) \equiv\left[z D_H(z) D_M^2(z)\right]^{1/3} \text {. }
\end{equation}

\begin{figure}[!htb]
   \begin{minipage}{0.49\textwidth}
     \centering
   \includegraphics[scale=0.35]{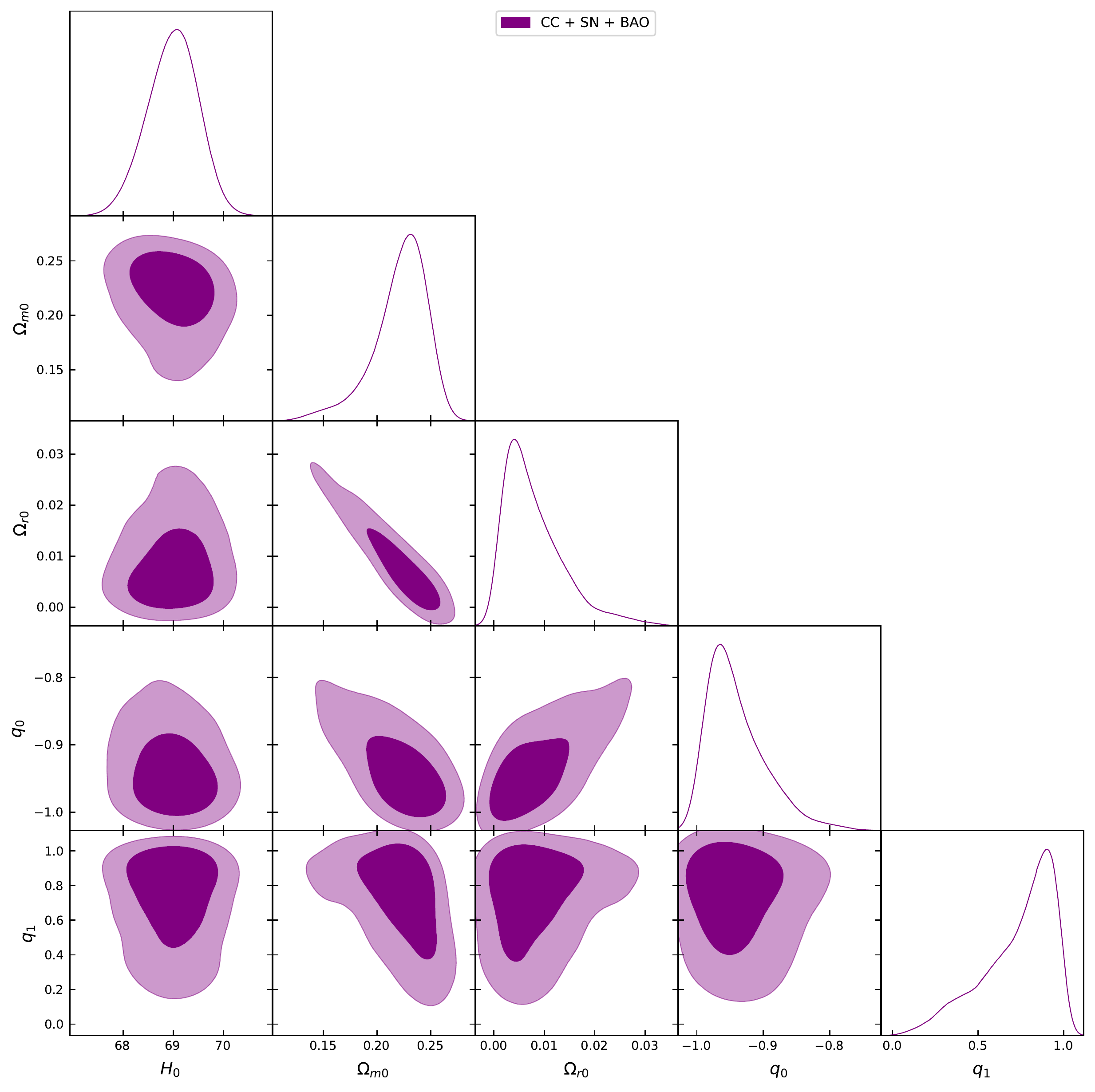}
\caption{The above figure shows the MCMC confidence contours at
1$\sigma$ and 2$\sigma$ obtained from CC+SNIa+BAO dataset for
Model 1 (Wetterich type).}\label{mcmc 1}
   \end{minipage}\hfill
   \begin{minipage}{0.49\textwidth}
     \centering
    \includegraphics[scale=0.35]{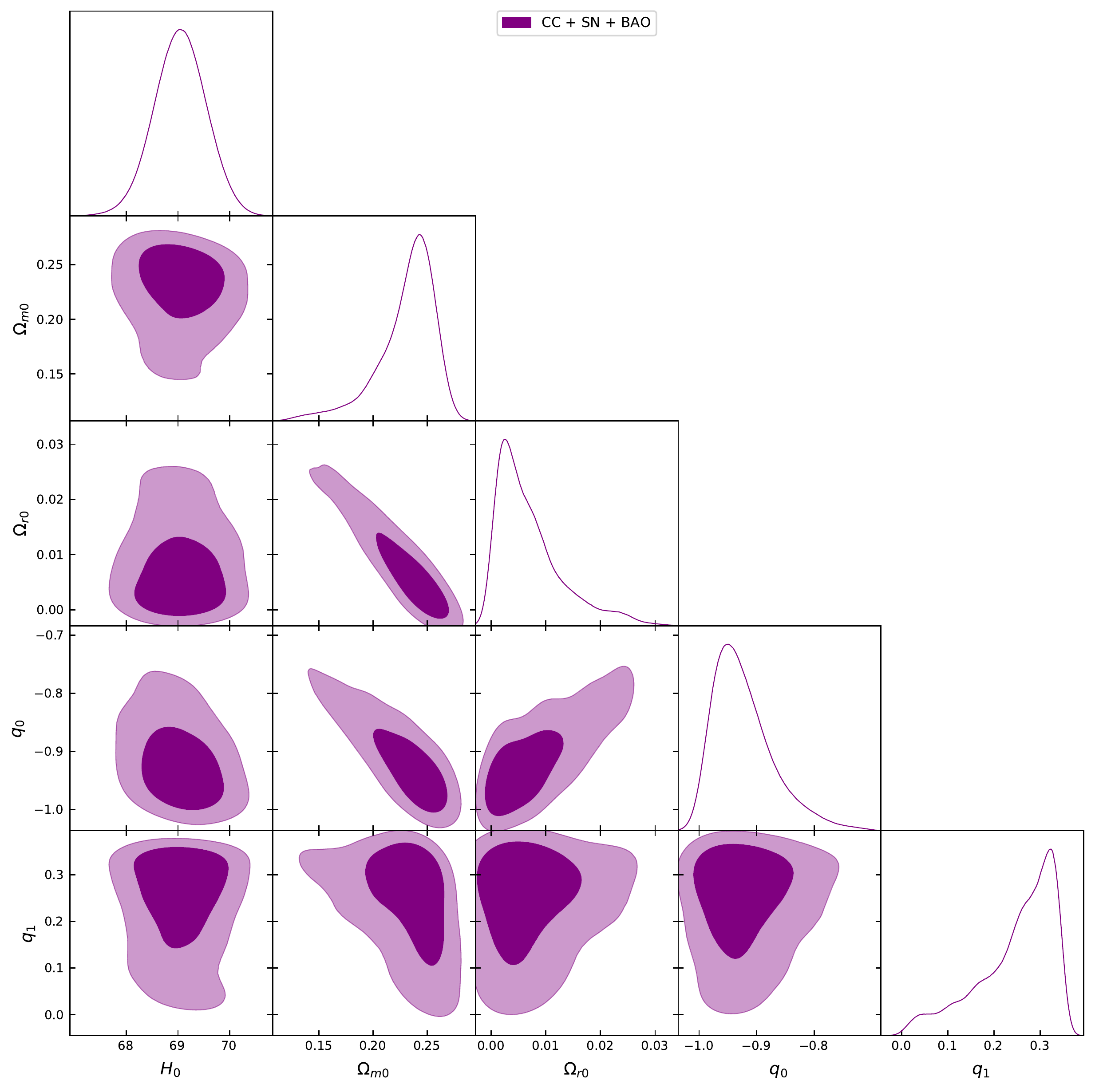}
\caption{The above figure shows the MCMC confidence contours at
1$\sigma$ and 2$\sigma$ obtained from CC+SNIa+BAO dataset for
Model 2 (Barboza-Alcaniz type).}\label{mcmc 2}
   \end{minipage}
\end{figure}
\begin{figure}[!htb]
   \begin{minipage}{0.49\textwidth}
     \centering
   \includegraphics[scale=0.35]{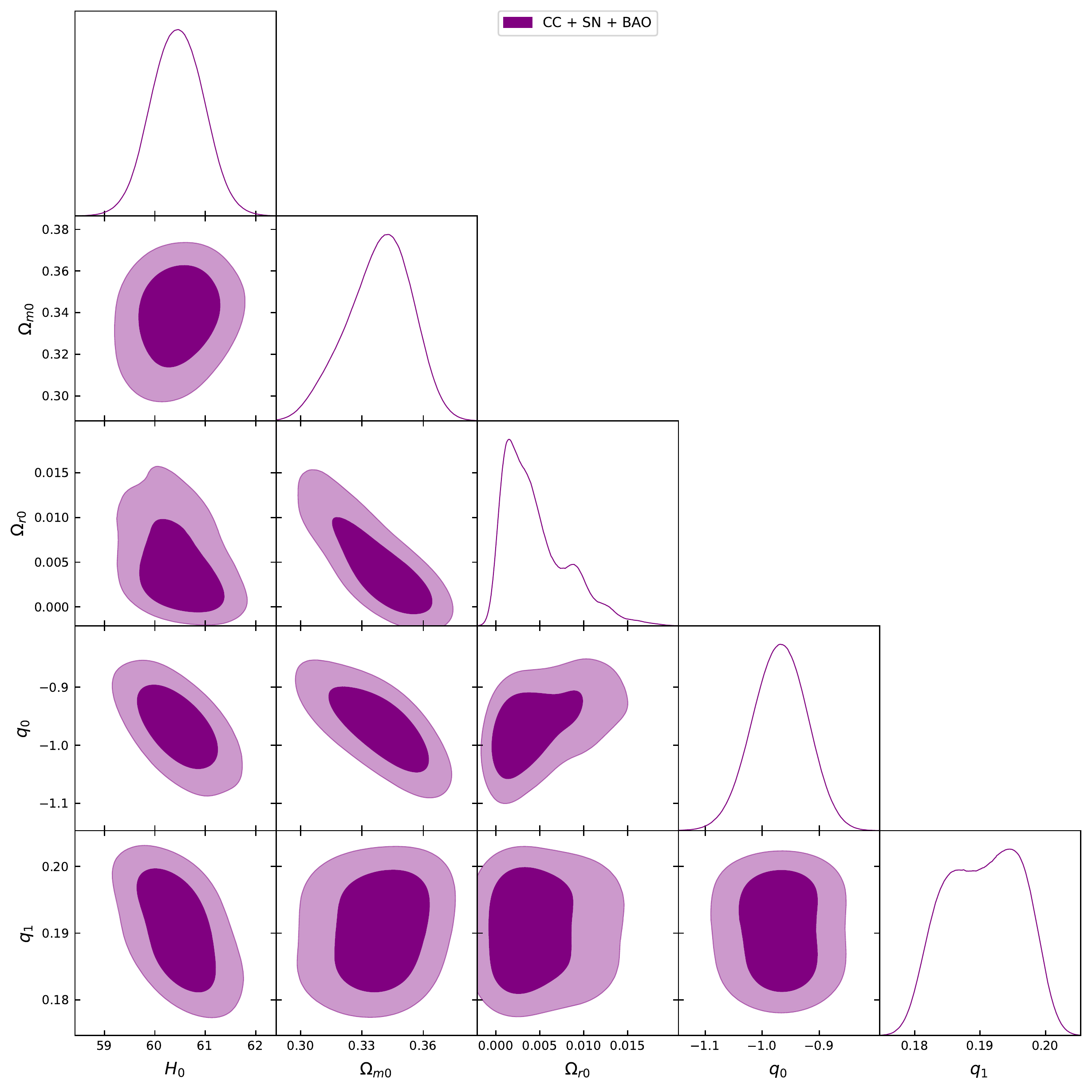}
\caption{The above figure shows the MCMC confidence contours at
1$\sigma$ and 2$\sigma$ obtained from CC+SNIa+BAO dataset for
Model 3 (CPL type).}\label{mcmc 3}
   \end{minipage}\hfill
   \begin{minipage}{0.49\textwidth}
     \centering
    \includegraphics[scale=0.35]{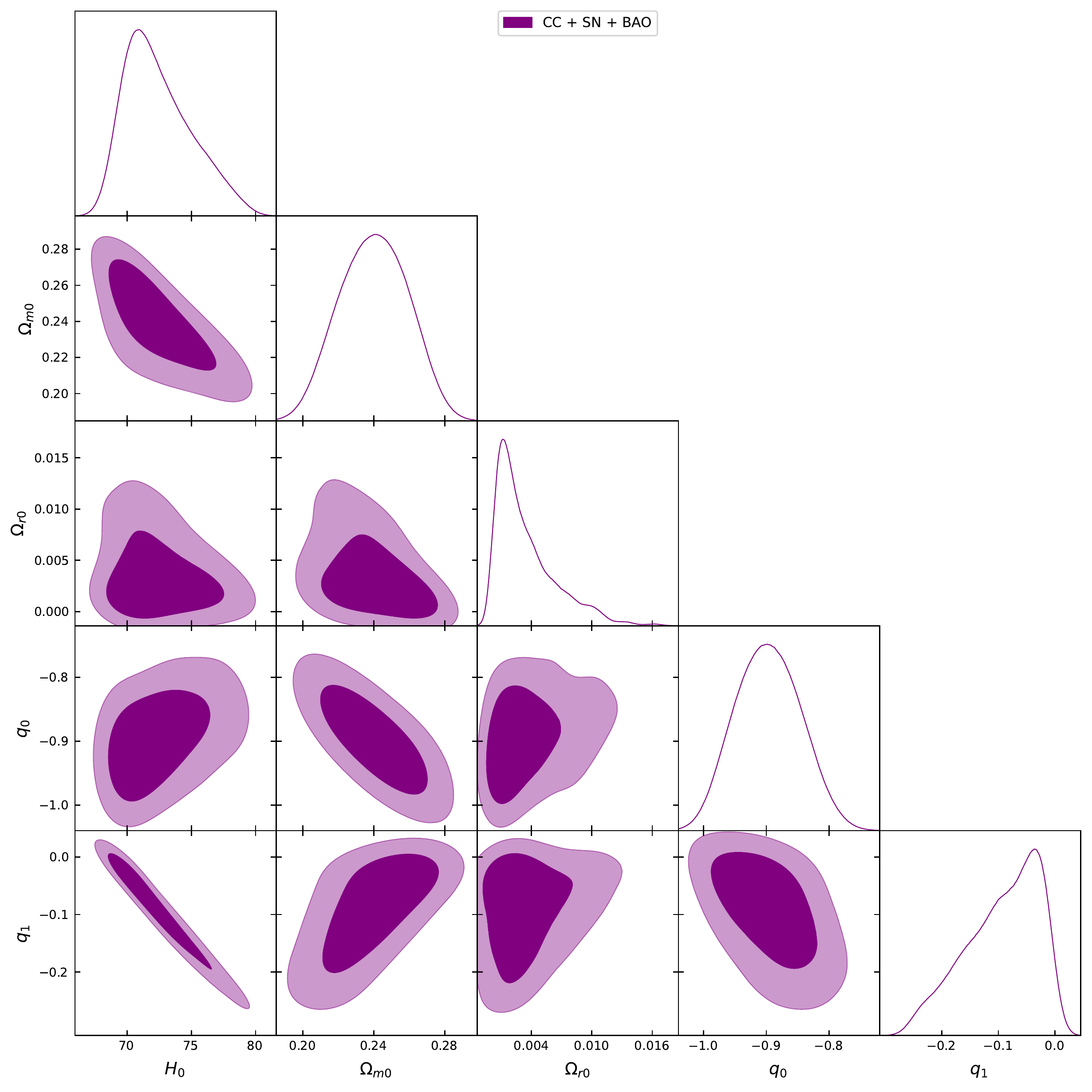}
\caption{The above figure shows the MCMC confidence contours at
1$\sigma$ and 2$\sigma$ obtained from CC+SNIa+BAO dataset for
Model 4.}\label{mcmc 4}
   \end{minipage}
\end{figure}

\begin{table}[H]
\begin{center}
\begin{tabular}{|c|c|c|c|}
\hline
\multicolumn{4}{|c|}{MCMC Results} \\
\hline
Model & Priors & Parameters & Best fit Value \\
\hline
$  \Lambda$CDM Model &$[50.,100.]$& $H_0$ &$69.854848_{-1.259100}^{+1.259100}$ \\
& $[0.,1.]$ &$\Omega_{\mathrm{m0}}$ & $0.268654_{-0.012822}^{+0.012822}$ \\
& $[0.,1]$ & $\Omega_\Lambda$ & $0.724585_{-0.009373}^{+0.009373}$ \\
\hline
Model 1 &  $[50.,100.]$ &$H_0$ & $68.990853_{-0.502428}^{+0.502428}$ \\
& $[0.,1.]$ &$\Omega_{\mathrm{m0}}$&  $0.221900_{-0.022982}^{+0.022982}$ \\
& $[0.,1]$ &$\Omega_{\mathrm{r0}}$&  $0.007835_{-0.005651}^{+0.005651}$ \\
& $[-1.5,-0.5]$ &$q_0$& $-0.941550_{-0.043352}^{+0.043352}$ \\
& $[0.,1.]$ &$q_1$& $0.740557_{-0.220903}^{+0.220903}$  \\
\hline
Model 2 & $[50.,100.]$ &$H_0$& $69.061167_{-0.498457}^{+0.498457}$ \\
& $[0.,1.]$ &$\Omega_{\mathrm{m0}}$&  $0.231719_{-0.025169}^{+0.025169}$ \\
& $[0.,1]$ &$\Omega_{\mathrm{r0}}$&  $0.006670_{-0.005154}^{+0.005154}$ \\
& $[-1.5,-0.5]$ &$q_0$& $-0.924723_{-0.052163}^{+0.052163}$ \\
& $[0.,1.]$ &$q_1$& $0.251461 _{-0.084597}^{+0.084597}$  \\
\hline
Model 3 & $[50.,100.]$ & $H_0$ &$60.450021_{-0.511464}^{+0.511464}$ \\
& $[0.,1.]$ &$\Omega_{\mathrm{m0}}$   &$0.338454_{-0.015533}^{+0.015533}$\\
& $[0.,1]$ &$\Omega_{\mathrm{r0}}$ & $0.004485_{-0.003527}^{+0.003527}$ \\
& $[-1.5,-0.5]$ &$q_0$ &$-0.968857_{-0.046034}^{+0.046034}$ \\
& $[0.,1.]$ &$q_1$ &$0.190554_{-0.006576}^{+0.006576}$ \\
\hline
Model 4 & $[50.,100.]$&$H_0$ &$71.392060 _{-2.608372}^{+2.608372}$ \\
& $[0.,1.]$ &$\Omega_{\mathrm{m0}}$   &$0.240129_{-0.021352}^{+0.021352}$\\
& $[0.,1]$ &$\Omega_{\mathrm{r0}}$ & $0.003402_{-0.002740}^{+0.002740}$ \\
& $[-1.5,-0.5]$&$q_0$ &$-0.897355_{-0.056271}^{+0.056271}$ \\
& $[-0.5,0.5]$ &$q_1$ &-$0.091362_{-0.073319}^{+0.073319 }$ \\
\hline
\end{tabular}
\caption{Summary of the MCMC results using CC + SNIa + BAO
dataset.}\label{tab_MCMC}
\end{center}
\end{table}

\newpage
\section{Observational, and theoretical comparisons of the Hubble and Distance Modulus functions}\label{sec5}
After determining the free parameters of Models 1-4, we can proceed to compare the predictions of these models with observational data. Additionally, we compare the model predictions to the well-established $\Lambda$CDM model, which serves as the base model for comparison. To provide a comprehensive analysis, we also consider the error bands associated with the $\Lambda$CDM model. This comparative analysis allows us to assess the viability and performance of the proposed models in capturing the observed data and to evaluate their agreement with the widely accepted $\Lambda$CDM model, providing insights into the potential strengths and limitations of each model in explaining the observed universe.

\subsection{Comparison with the Hubble data points.}
\begin{figure}[!htb]
   \begin{minipage}{0.49\textwidth}
     \centering
   \includegraphics[scale=0.58]{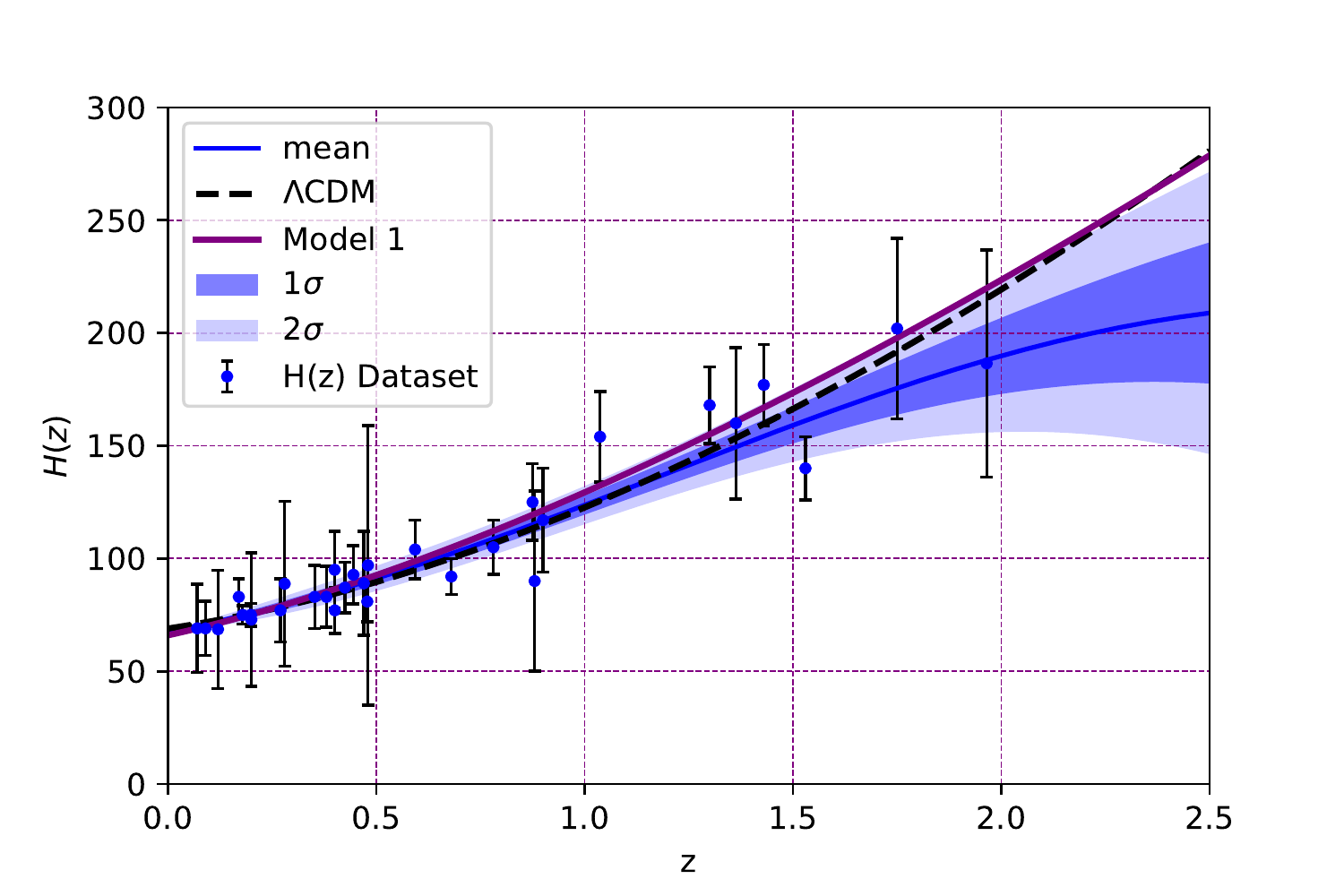}
\caption{The figure shows that the theoretical curve of the Hubble function $H(z)$ of Model 1 is shown in purple line and $\Lambda$CDM Model shown in black dotted line with $\Omega_{\mathrm{m0}}=$ 0.3 and $\Omega_\Lambda =$ 0.7 against 57 $H(z)$ datasets are shown in blue dots with their corresponding error bars with 1$\protect\sigma$ and 2$\protect\sigma$ error bands.}\label{CC Model 1}
   \end{minipage}
\end{figure}

\begin{figure}[!htb]
   \begin{minipage}{0.49\textwidth}
     \centering
    \includegraphics[scale=0.58]{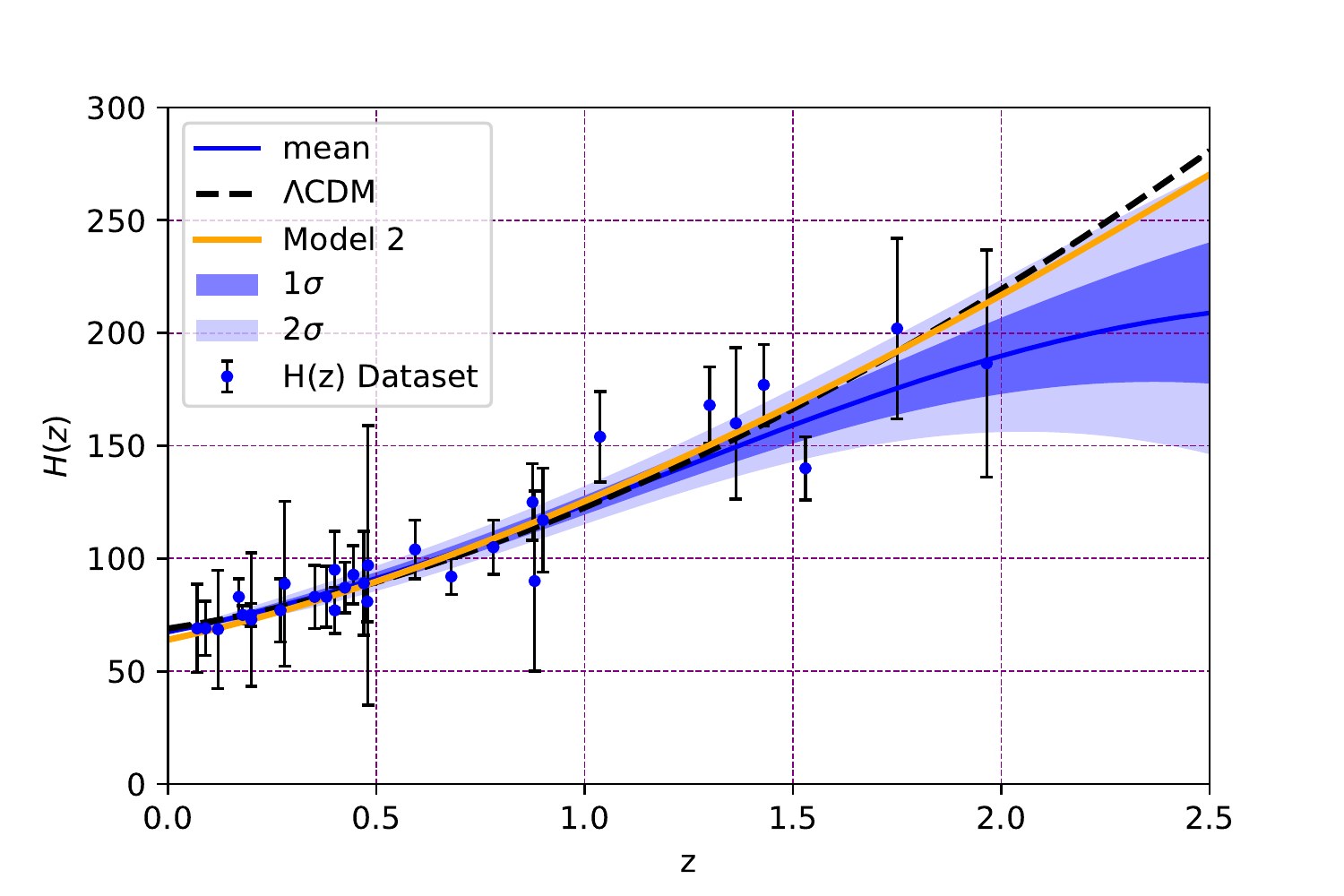}
\caption{The figure shows that the theoretical curve of the Hubble function $H(z)$ of Model 2 is shown in orange line and $\Lambda$CDM Model shown in black dotted line with $\Omega_{\mathrm{m0}}=$ 0.3 and $\Omega_\Lambda =$ 0.7 against 57 $H(z)$ datasets are shown in blue dots with their corresponding error bars with 1$\protect\sigma$ and 2$\protect\sigma$ error bands.}\label{CC Model 2}
   \end{minipage}\hfill
   \begin{minipage}{0.49\textwidth}
     \centering
   \includegraphics[scale=0.58]{fig_6.pdf}
\caption{The figure shows that the theoretical curve of the Hubble function $H(z)$ of Model 3 is shown in green line and $\Lambda$CDM Model shown in black dotted line with $\Omega_{\mathrm{m0}}=$ 0.3 and $\Omega_\Lambda =$ 0.7 against 57 $H(z)$ datasets are shown in blue dots with their corresponding error bars with 1$\protect\sigma$ and 2$\protect\sigma$ error bands.}\label{CC Model 3}
   \end{minipage}\hfill
   \begin{minipage}{0.49\textwidth}
     \centering
   \includegraphics[scale=0.6]{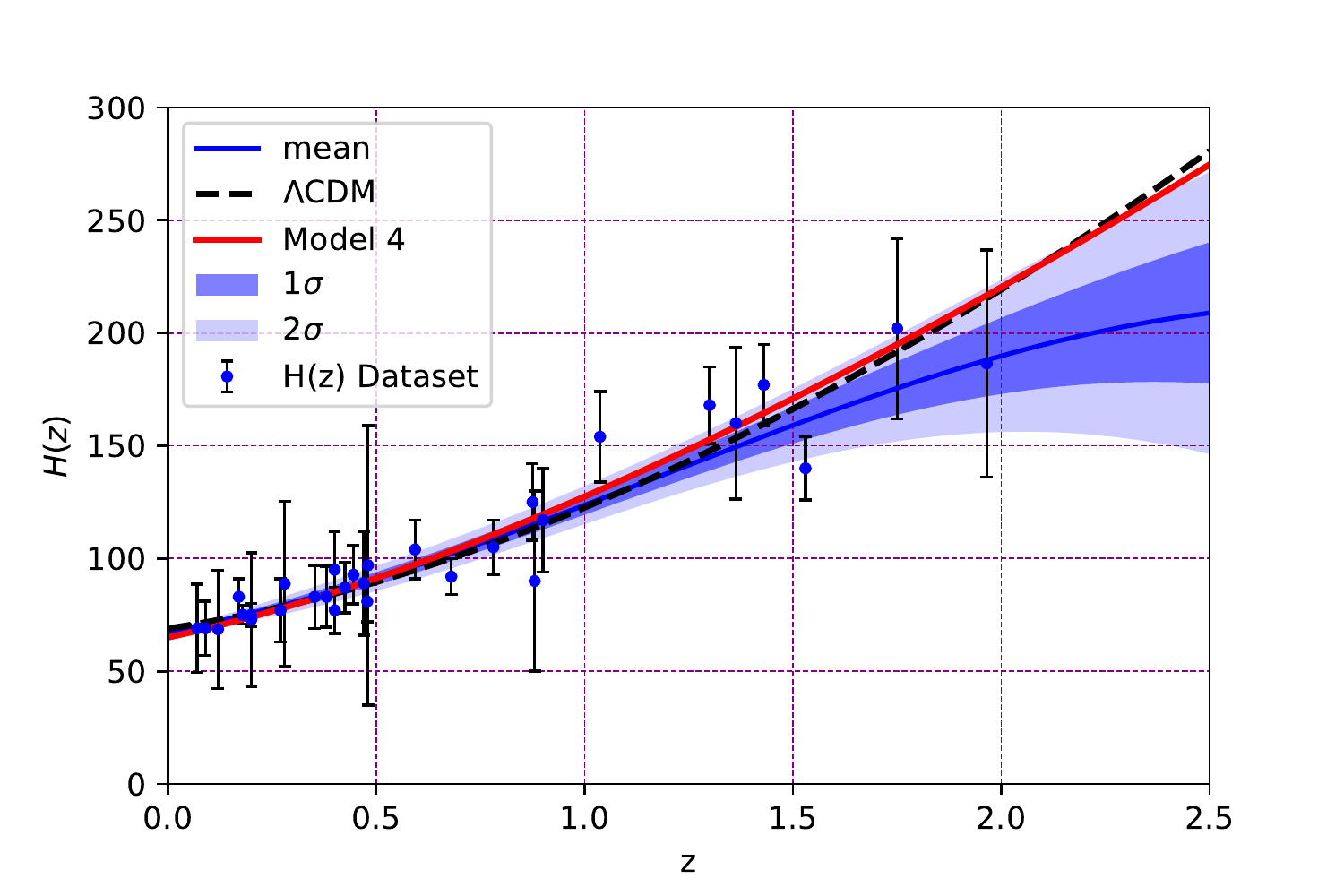}
\caption{The figure shows that the theoretical curve of the Hubble function $H(z)$ of Model 4 is shown in red line and $\Lambda$CDM Model shown in black dotted line with $\Omega_{\mathrm{m0}}=$ 0.3 and $\Omega_\Lambda =$ 0.7 against 57 $H(z)$ datasets are shown in blue dots with their corresponding error bars with 1$\protect\sigma$ and 2$\protect\sigma$ error bands.}\label{CC Model 4}
   \end{minipage}
\end{figure}

First, We consider the comparison of the Models (1 - 4) with the
31 (CC) data points and the $\Lambda$CDM model with 1 $\sigma$ and 2 $\sigma$ error bands . The comparison findings are shown in
Figure \ref{CC Model 1}, \ref{CC Model 2}, \ref{CC Model 3} and
\ref{CC Model 4}. The Figure shows that all models fit with (CC)
dataset quite well.

\subsection{Comparison with the type Ia supernova dataset.} 
\begin{figure}[!htb]
   \begin{minipage}{0.49\textwidth}
     \centering
   \includegraphics[scale=0.6]{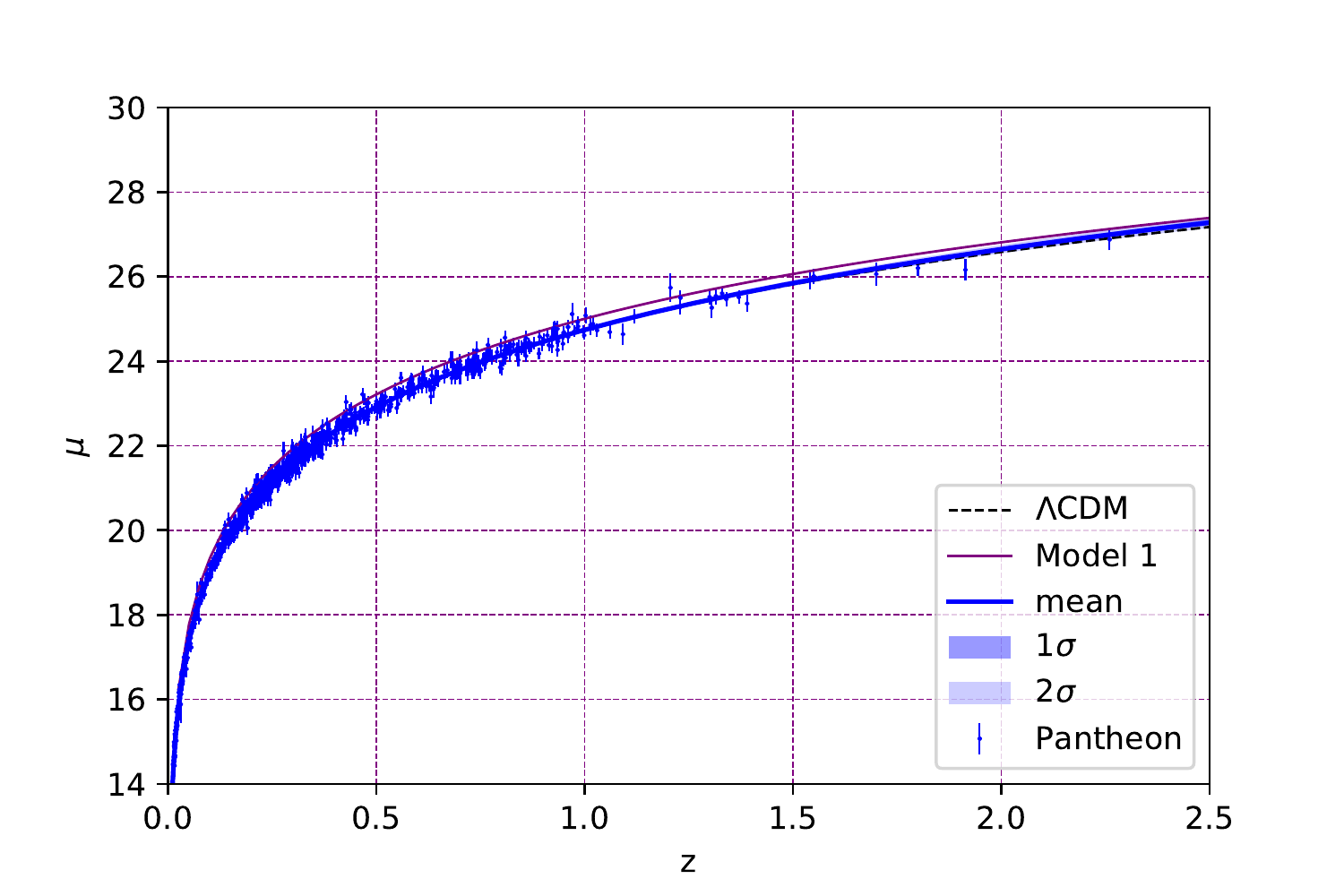}
\caption{Theoretical curve of distance modulus $\protect\mu(z) $ of the Model 1 is shown in purple line and the $\Lambda$CDM Model is shown in the black dotted line with $\Omega_{\mathrm{m0}}=$ 0.3 and $\Omega_\Lambda =$ 0.7 against type Ia supernova data are shown in blue dots with their corresponding errors bars with 1$\protect\sigma$ and 2$\protect\sigma$ error bands.}\label{Pan Model 1}
   \end{minipage}\hfill
   \begin{minipage}{0.49\textwidth}
     \centering
   \includegraphics[scale=0.58]{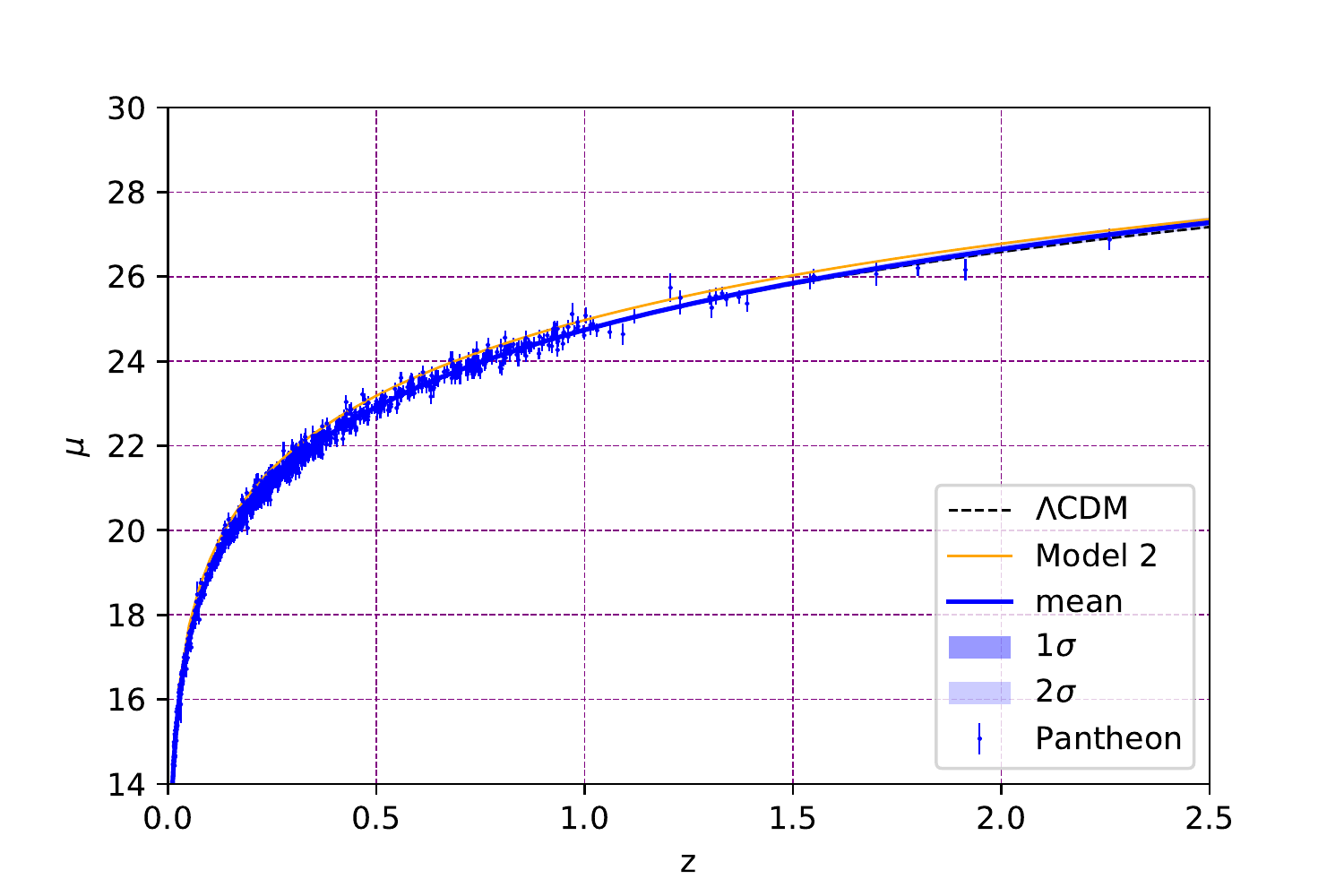}
\caption{Theoretical curve of distance modulus $\protect\mu(z) $ of the Model 2 is shown in orange line and the $\Lambda$CDM Model is shown in the black dotted line with $\Omega_{\mathrm{m0}}=$ 0.3 and $\Omega_\Lambda =$ 0.7, against type Ia supernova data are shown in blue dots with their corresponding errors bars with 1$\protect\sigma$ and 2$\protect\sigma$ error bands.}\label{Pan Model 2}
   \end{minipage}
\end{figure}

\begin{figure}[!htb]
   \begin{minipage}{0.49\textwidth}
     \centering
    \includegraphics[scale=0.58]{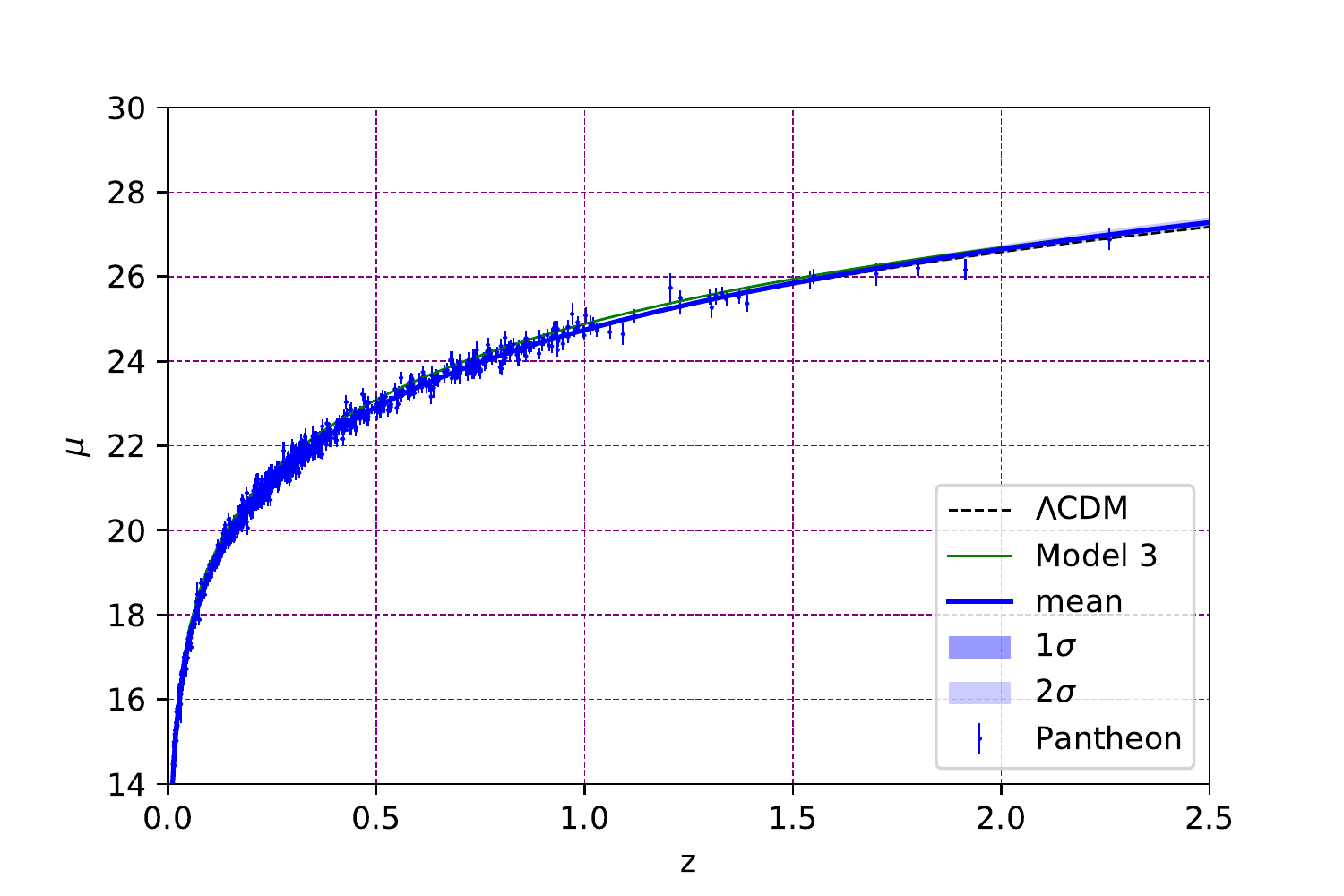}
\caption{Theoretical curve of distance modulus $\protect\mu(z) $ of the Model 3 is shown in green line and the $\Lambda$CDM Model is shown in the black dotted line with $\Omega_{\mathrm{m0}}=$ 0.3 and $\Omega_\Lambda =$ 0.7, against type Ia supernova data are shown in blue dots with their corresponding errors bars with 1$\protect\sigma$ and 2$\protect\sigma$ error bands.}\label{Pan Model 3}
   \end{minipage}\hfill
   \begin{minipage}{0.49\textwidth}
     \centering
    \includegraphics[scale=0.58]{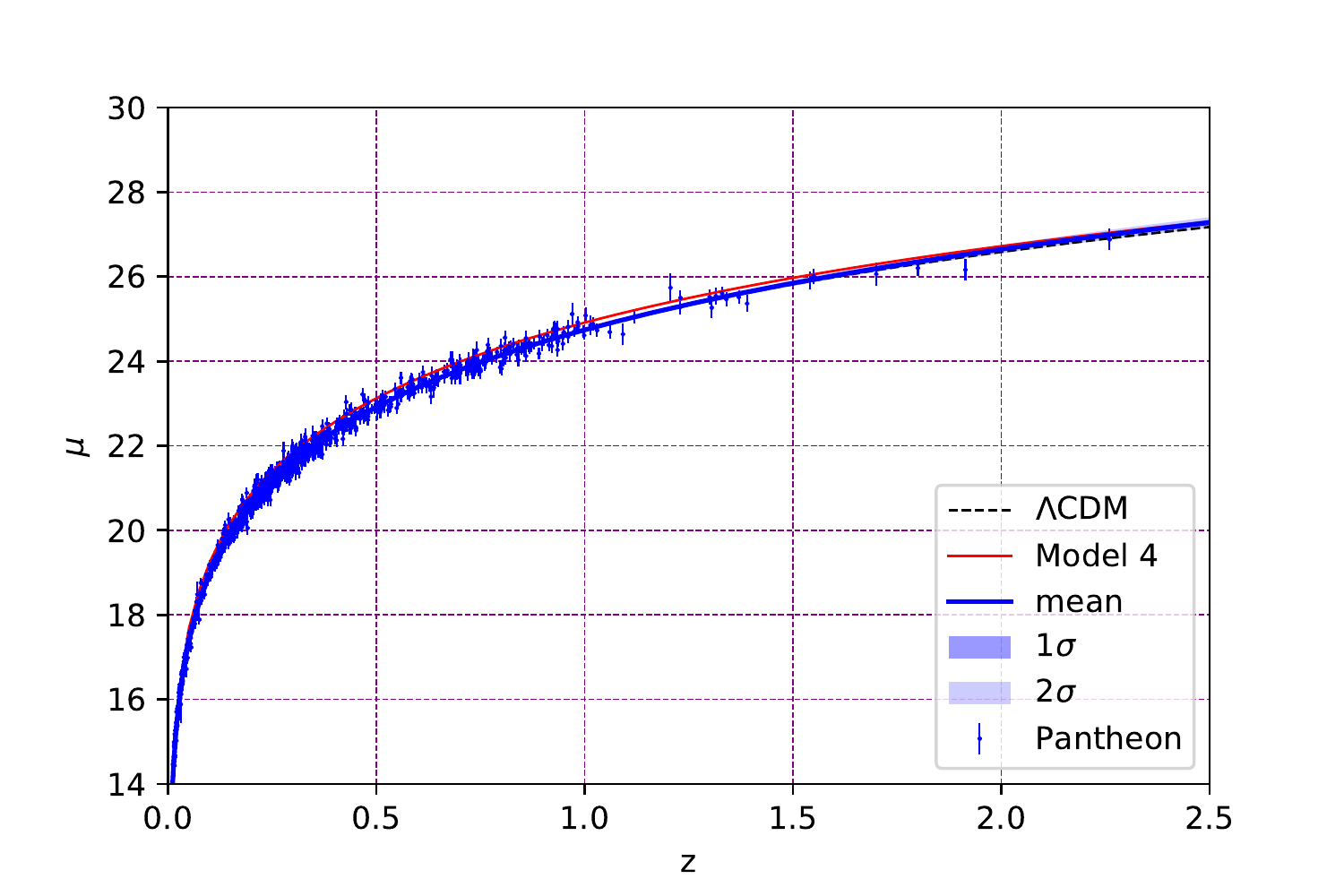}
\caption{Theoretical curve of distance modulus $\protect\mu(z) $ of the Model 4 is shown in red line and the $\Lambda$CDM Model is shown in the black dotted line with $\Omega_{\mathrm{m0}}=$ 0.3 and $\Omega_\Lambda =$ 0.7, against type Ia supernova data are shown in blue dots with their corresponding errors bars with 1$\protect\sigma$ and 2$\protect\sigma$ error bands.}\label{Pan Model 4}
   \end{minipage}
\end{figure}

We now compare the $\mu(z)$ distance modulus function of Models (1-4) with the type Ia supernova dataset. From Fig.~\ref{Pan Model 1},\ref{Pan Model 2},\ref{Pan Model 3} and \ref{Pan Model 4} one can see that all Models fit with the type Ia supernova dataset, very well.

\clearpage
\section{Cosmography Parameters}\label{sec6}
To study the early evolution and late evolution of the universe,
some other parameters named cosmographical parameters can be
analyzed. The cosmographical parameter like jerk ($j$), snap ($s$)
parameters are \cite{48,49,50,51}

\begin{eqnarray}
j=\frac{\dddot{a}}{aH^{3}}=(1+z)\frac{dq}{dz}+q(1+2q),\\
s=\frac{a^{(4)}}{aH^{4}}=-(1+z)\frac{dj}{dz}-j(2+3q)\\
\end{eqnarray}

So the cosmographical parameters contain the higher-order derivatives of the deceleration parameter $q$. The 'jerk' parameter is considered to have a very useful feature that is for standard $\Lambda$CDM model, $j$ always takes the value unity which helps us assess the deviation regarding different dark energy models. Sahni et al. and Alam et al. analyzed the importance of the jerk parameter $j$ for discriminating various dark energy models. We have explored the evolution of such kinematical quantities with respect to the redshift for the involved parametric models.\\

\begin{figure}[!htb]
   \begin{minipage}{0.49\textwidth}
     \centering
   \includegraphics[scale=0.4]{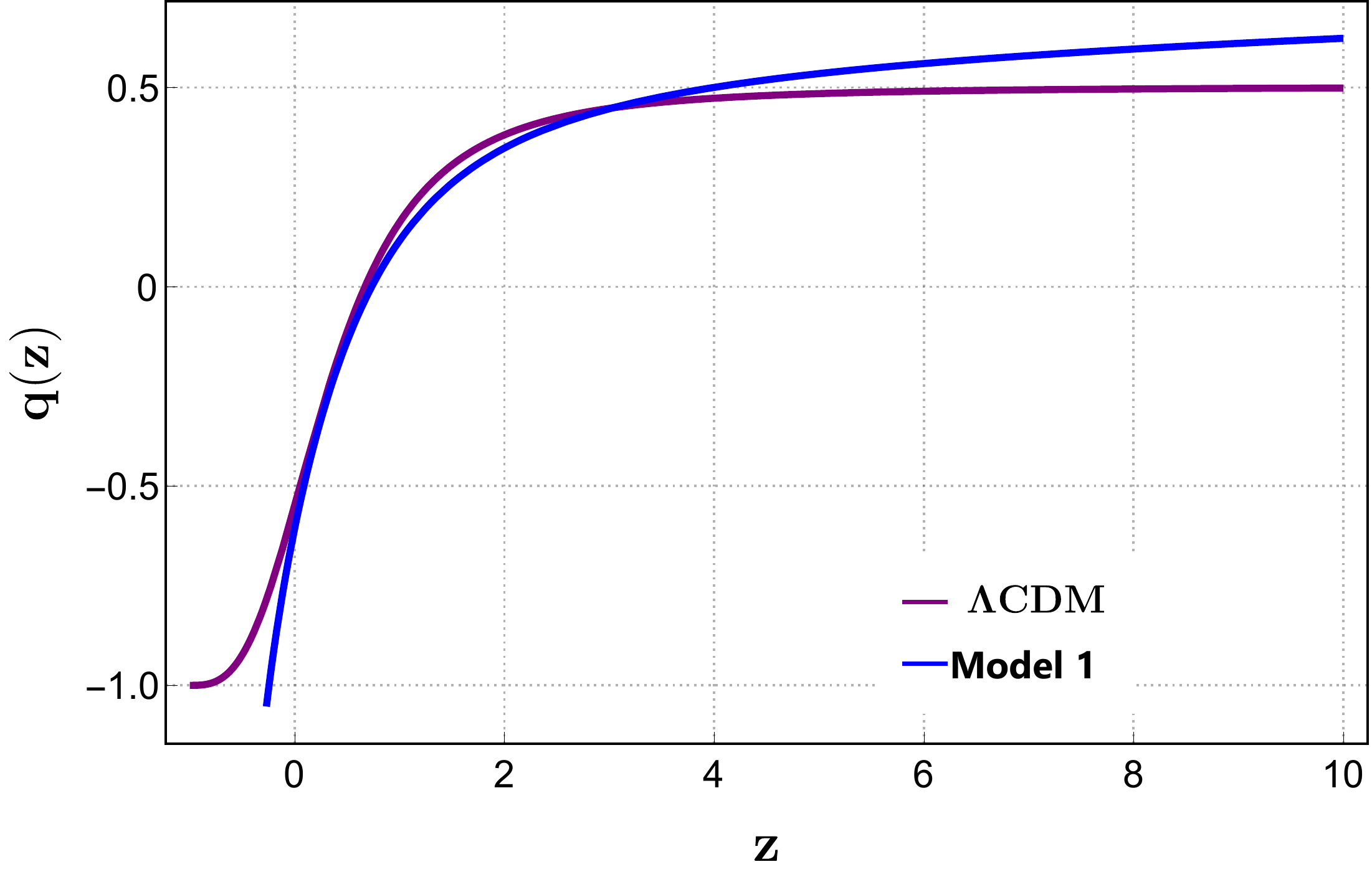}
\caption{Evolution of deceleration parameter of Model 1  with
respect to redshift.}\label{q(z) Model 1}
   \end{minipage}
\end{figure}
\begin{figure}[!htb]
   \begin{minipage}{0.49\textwidth}
     \centering
   \includegraphics[scale=0.4]{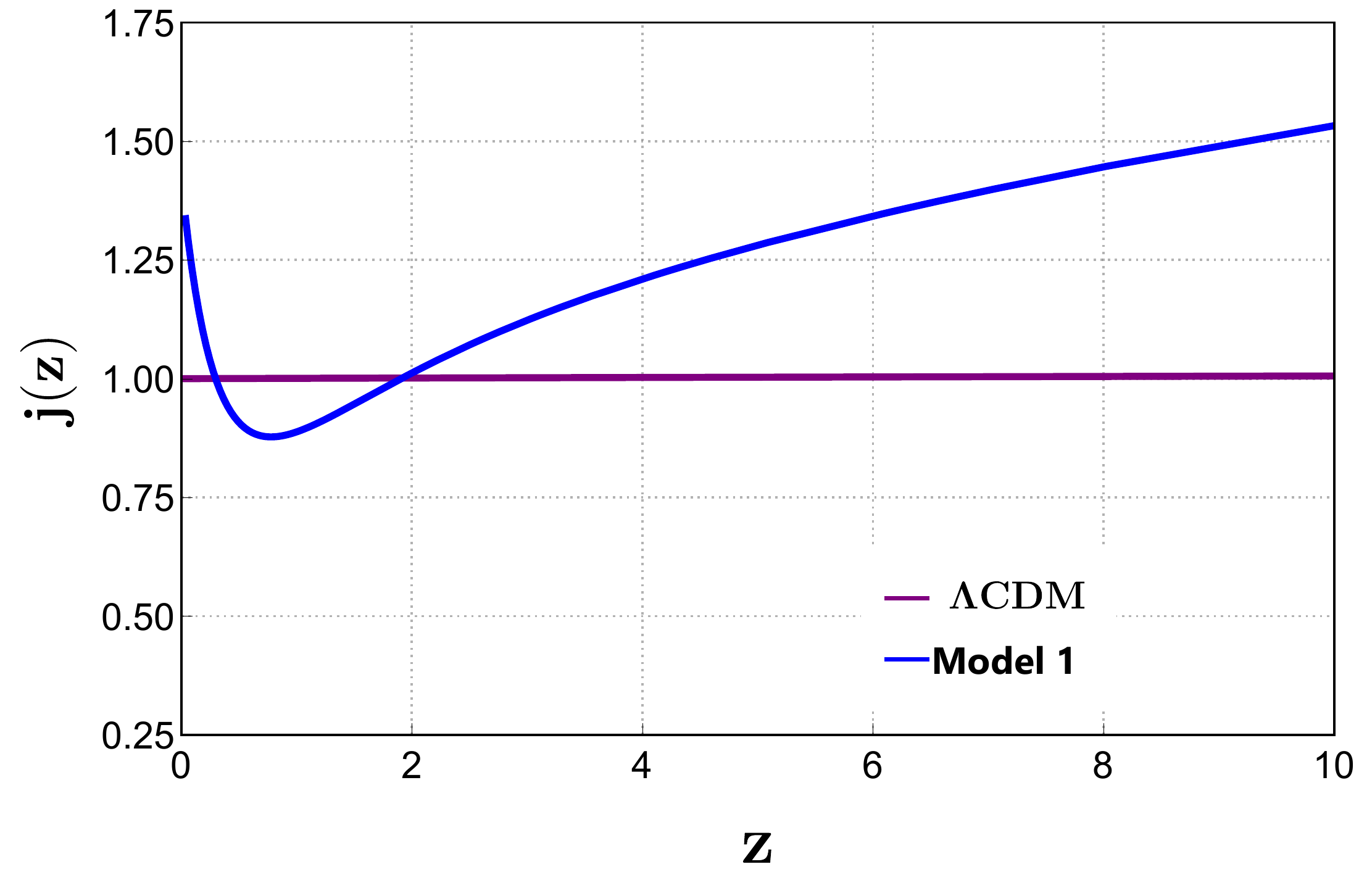}
\caption{Evolution of jerk parameter of Model 2  with respect to
redshift.}\label{j(z) Model 1}
   \end{minipage}\hfill
   \begin{minipage}{0.49\textwidth}
     \centering
    \includegraphics[scale=0.39]{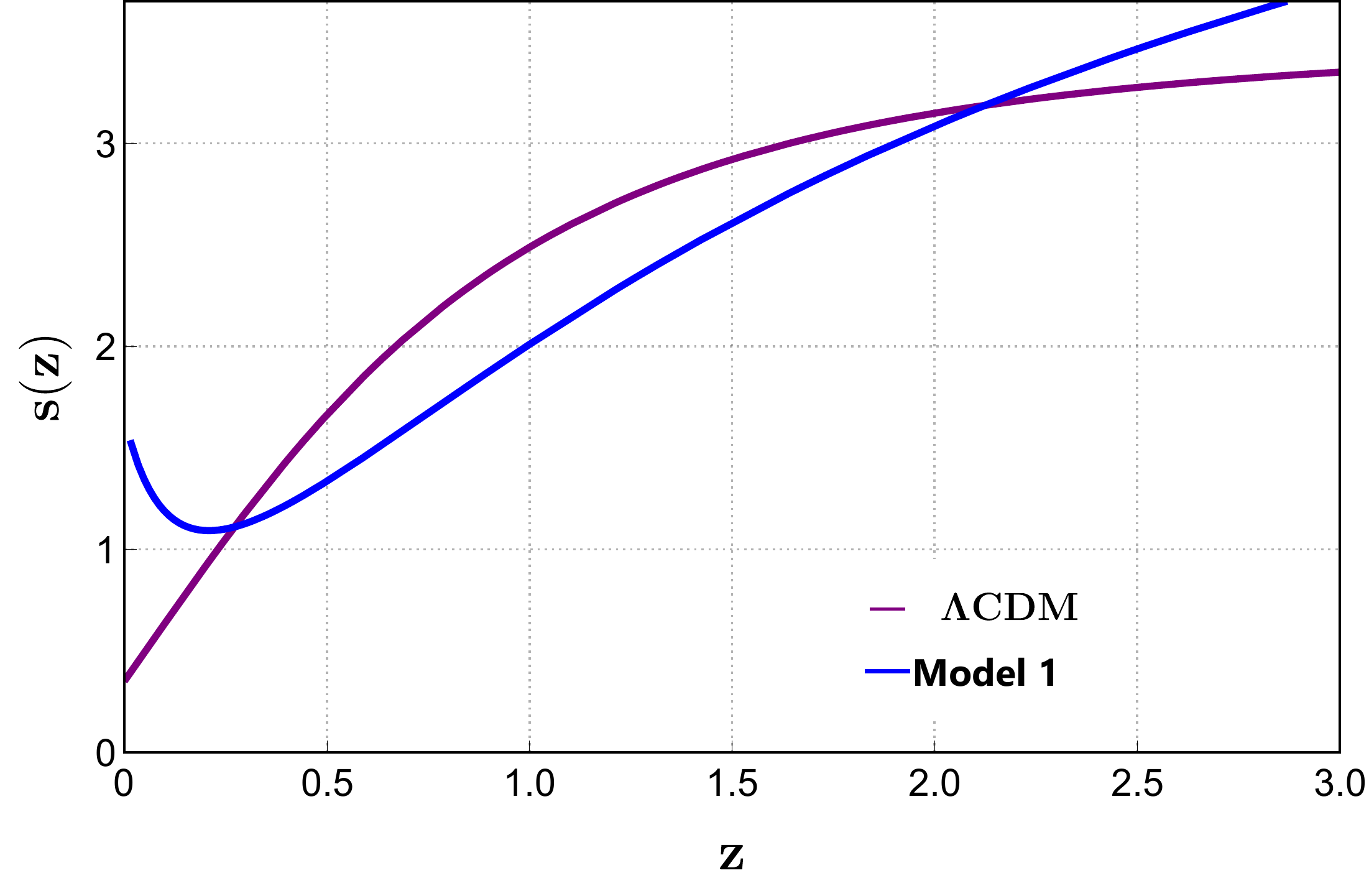}
\caption{Evolution of snap parameter with of Model 3 with respect
to redshift.}\label{s(z) Model 1}
   \end{minipage}\hfill
   \begin{minipage}{0.49\textwidth}
     \centering
    \includegraphics[scale=0.39]{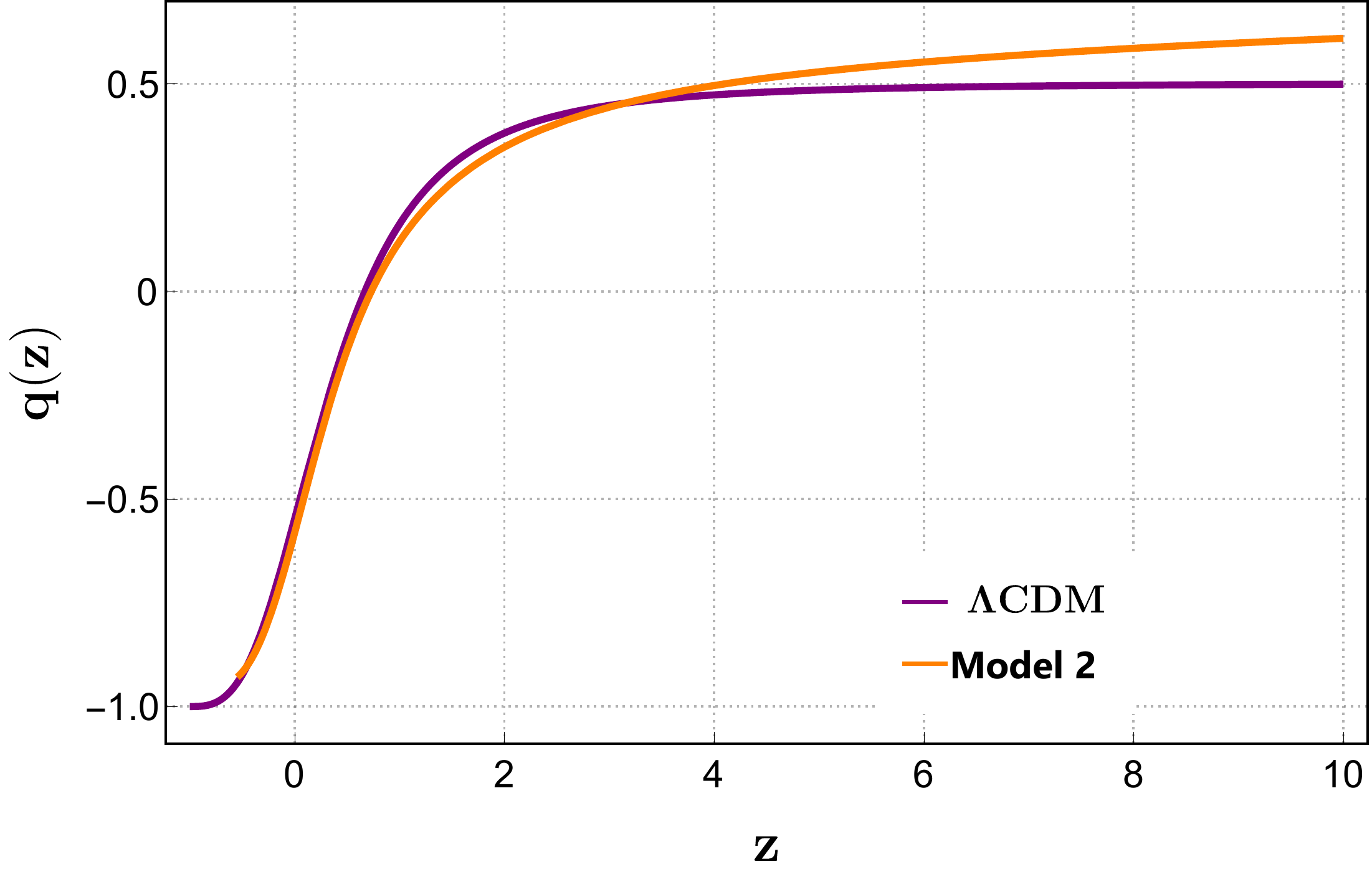}
\caption{Evolution of deceleration parameter of Model 1  with
respect to redshift.}\label{q(z) Model 2}
   \end{minipage}
\end{figure}

\begin{figure}[!htb]
   \begin{minipage}{0.49\textwidth}
     \centering
   \includegraphics[scale=0.4]{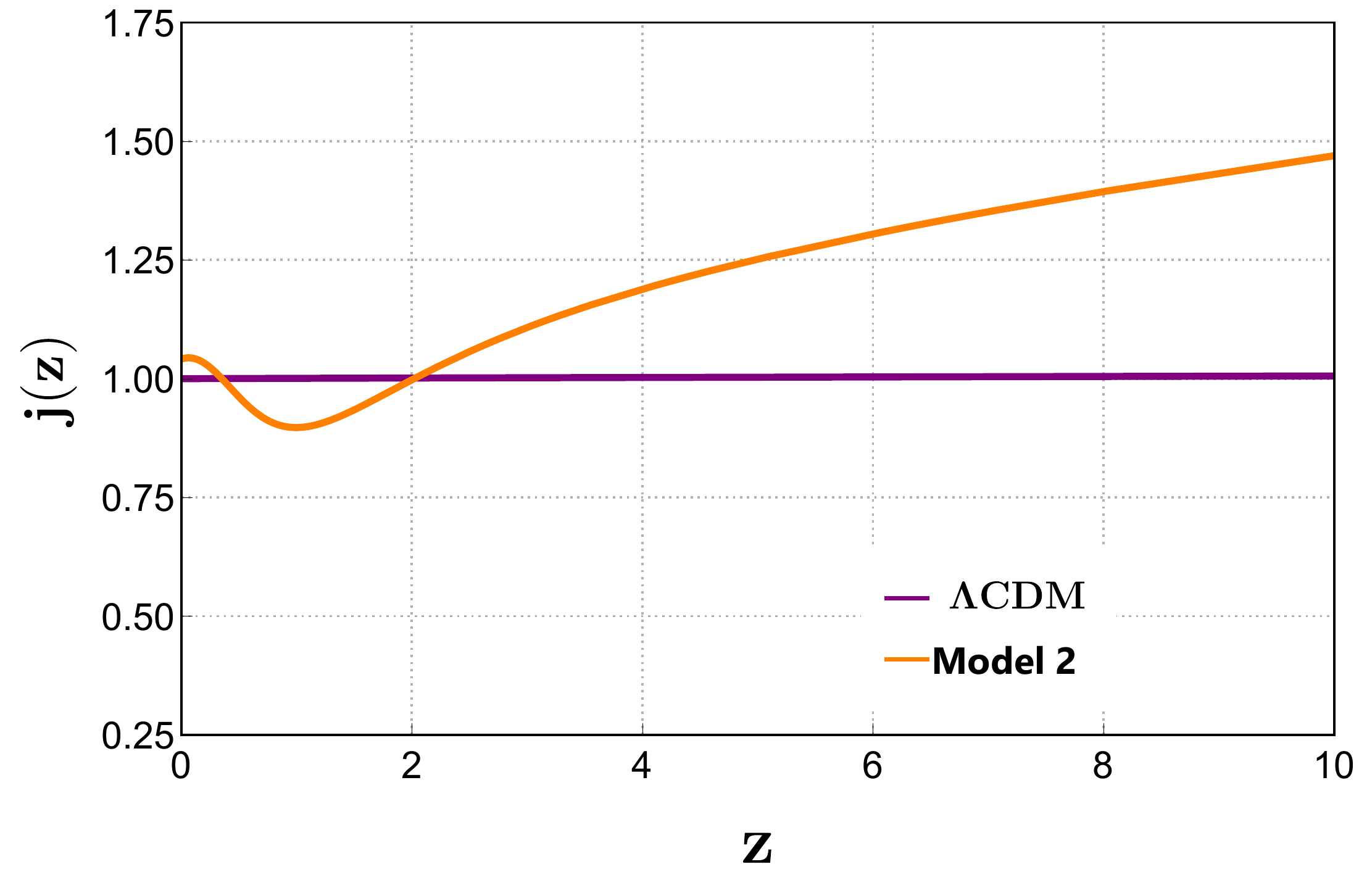}
\caption{Evolution of jerk parameter of Model 2  with respect to
redshift.}\label{j(z) Model 2}
   \end{minipage}\hfill
   \begin{minipage}{0.49\textwidth}
     \centering
    \includegraphics[scale=0.4]{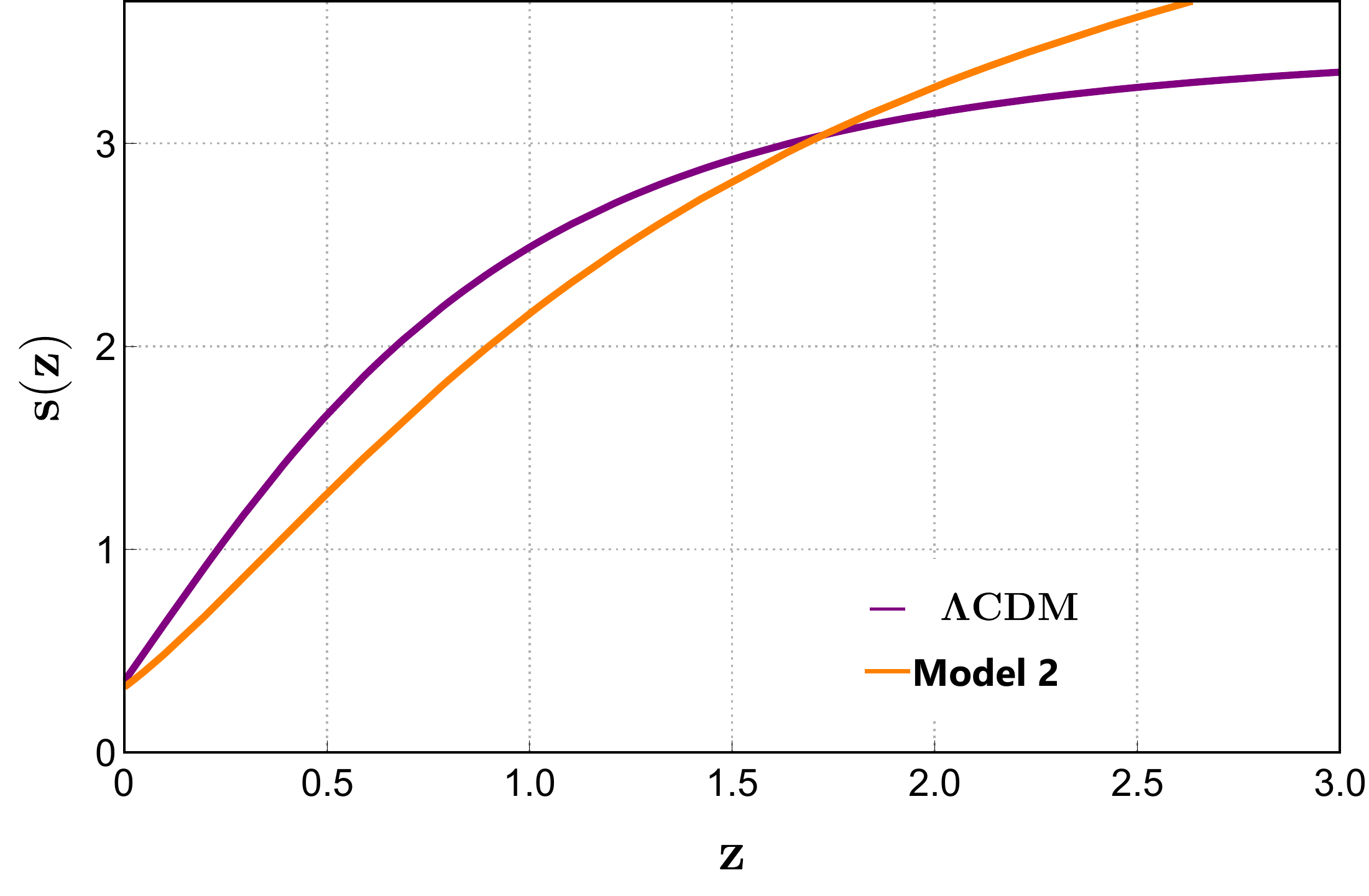}
\caption{Evolution of snap parameter with of Model 2 with respect
to redshift.}\label{s(z) Model 2}
   \end{minipage}\hfill
   \begin{minipage}{0.49\textwidth}
     \centering
    \includegraphics[scale=0.4]{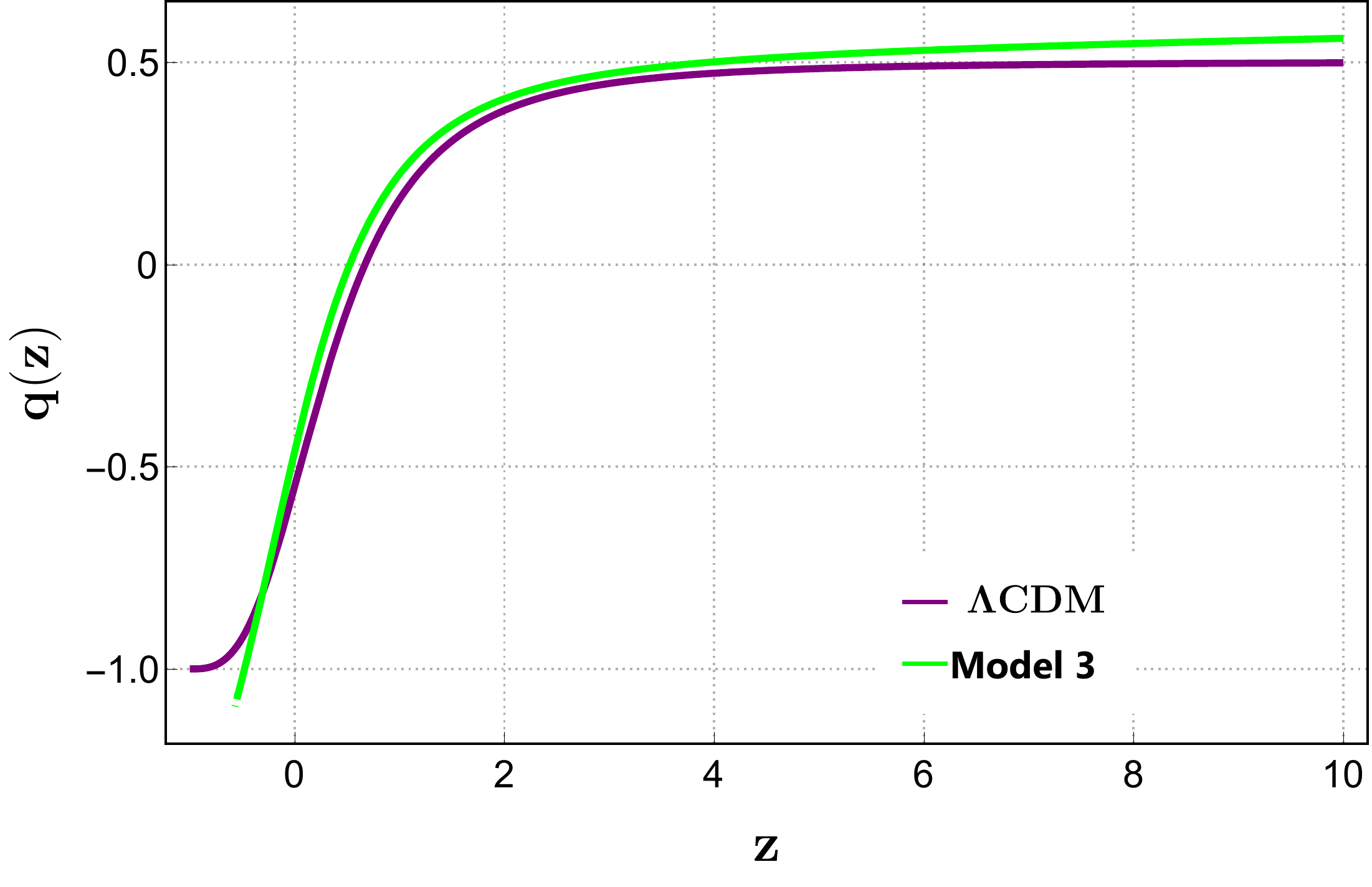}
\caption{Evolution of decceleration parameter with of Model 3 with
respect to redshift.}\label{q(z) Model 3}
   \end{minipage}
\end{figure}

\begin{figure}[!htb]
   \begin{minipage}{0.49\textwidth}
     \centering
   \includegraphics[scale=0.38]{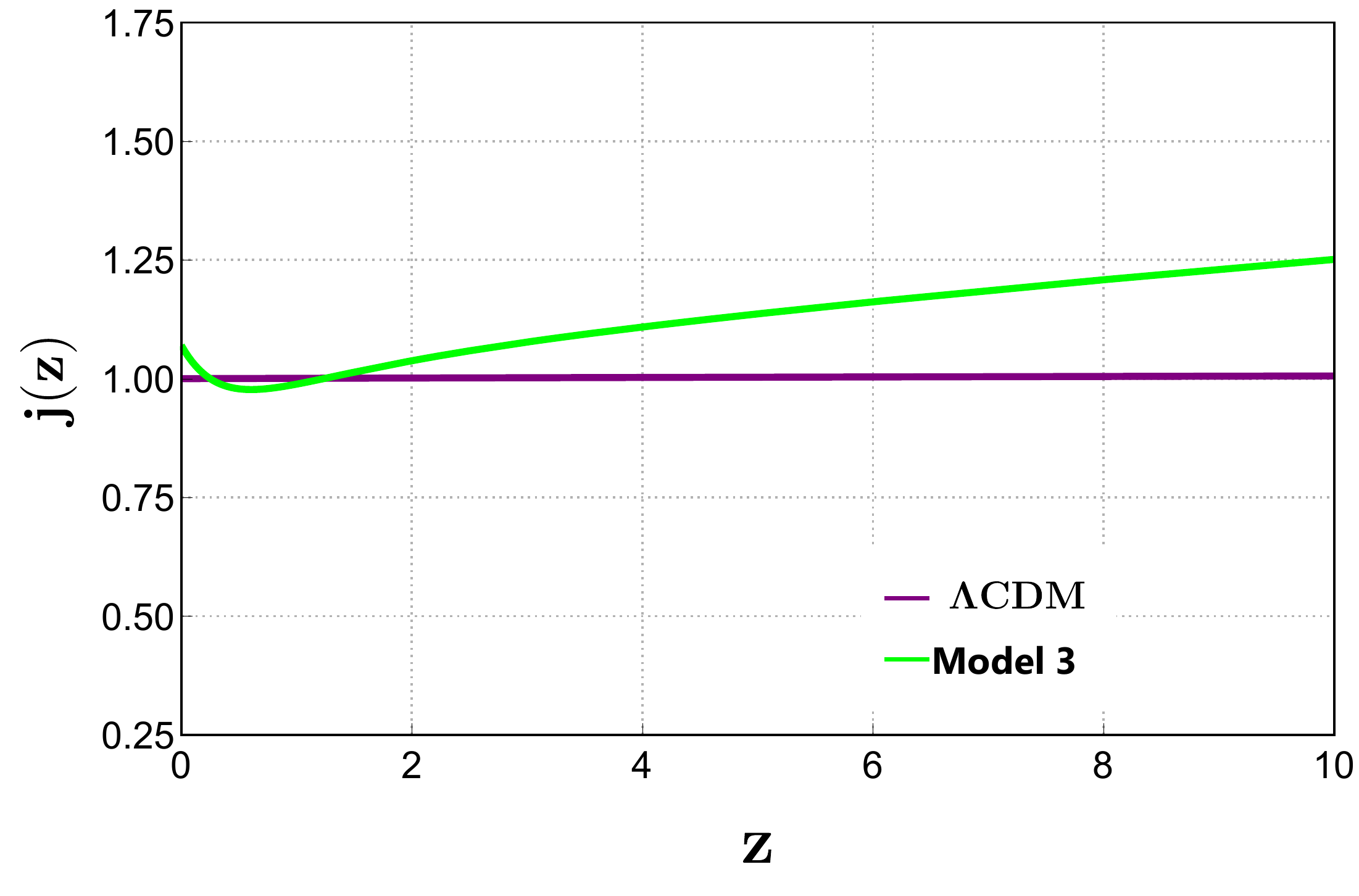}
\caption{Evolution of jerk parameter of Model 2  with respect to
redshift.}\label{j(z) Model 3}
   \end{minipage}\hfill
   \begin{minipage}{0.49\textwidth}
     \centering
    \includegraphics[scale=0.38]{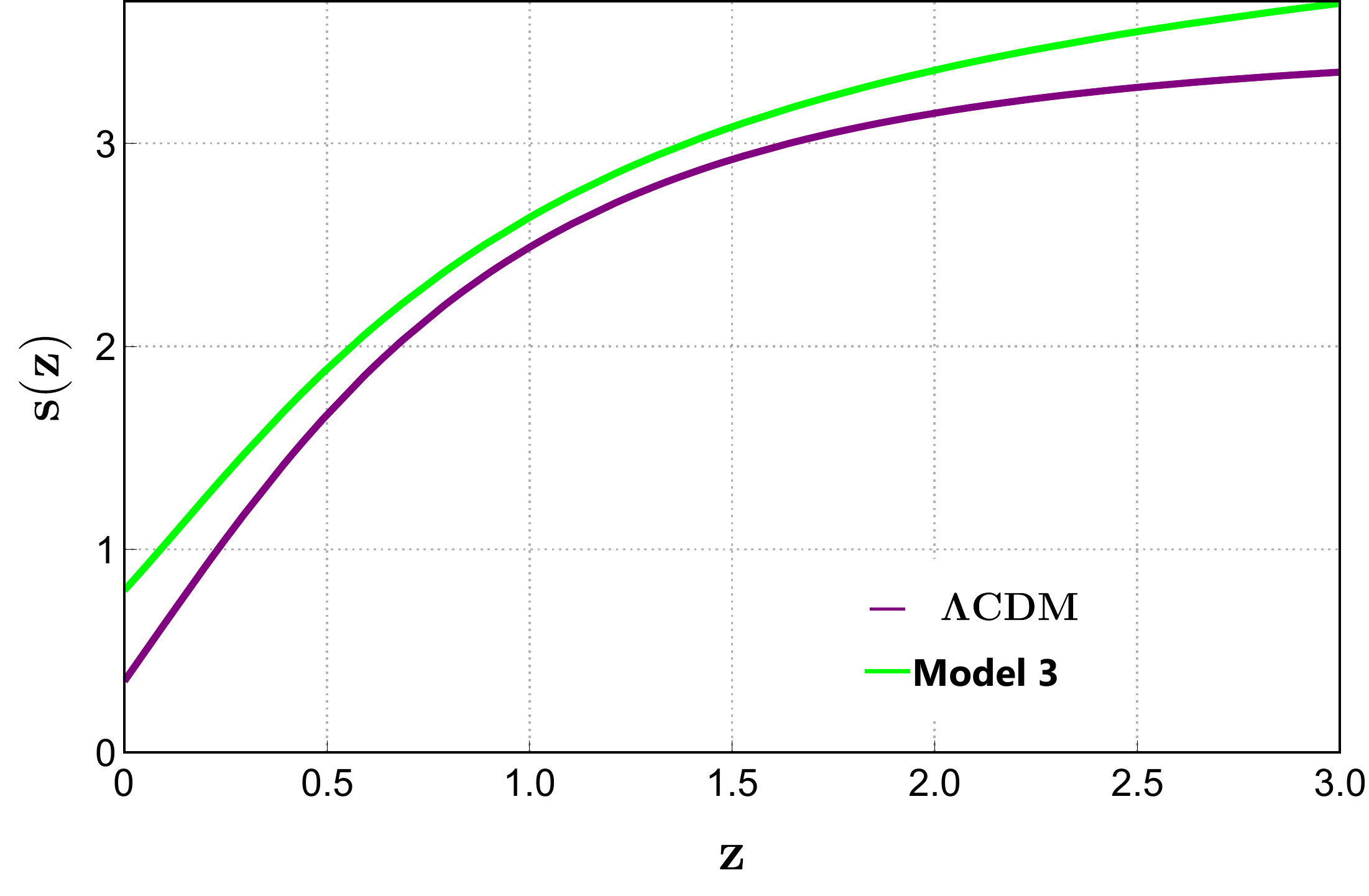}
\caption{Evolution of snap parameter with of Model 3 with respect
to redshift.}\label{s(z) Model 3}
   \end{minipage}\hfill
   \begin{minipage}{0.49\textwidth}
     \centering
    \includegraphics[scale=0.42]{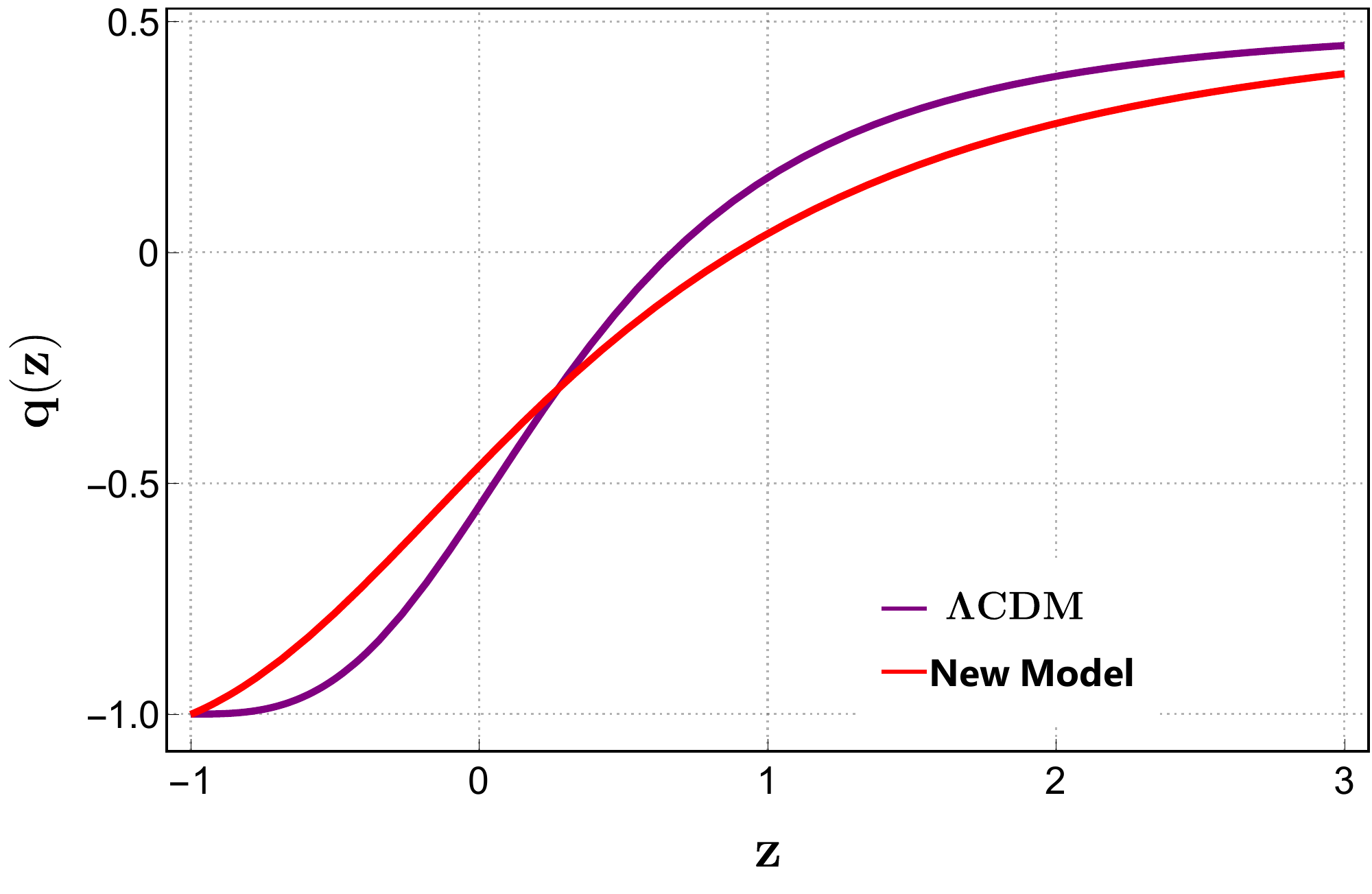}
\caption{Evolution of deceleration parameter with of Model 4 with
respect to redshift.}\label{q(z) Model 4}
   \end{minipage}
\end{figure}

\begin{figure}[!htb]
   \begin{minipage}{0.49\textwidth}
     \centering
   \includegraphics[scale=0.42]{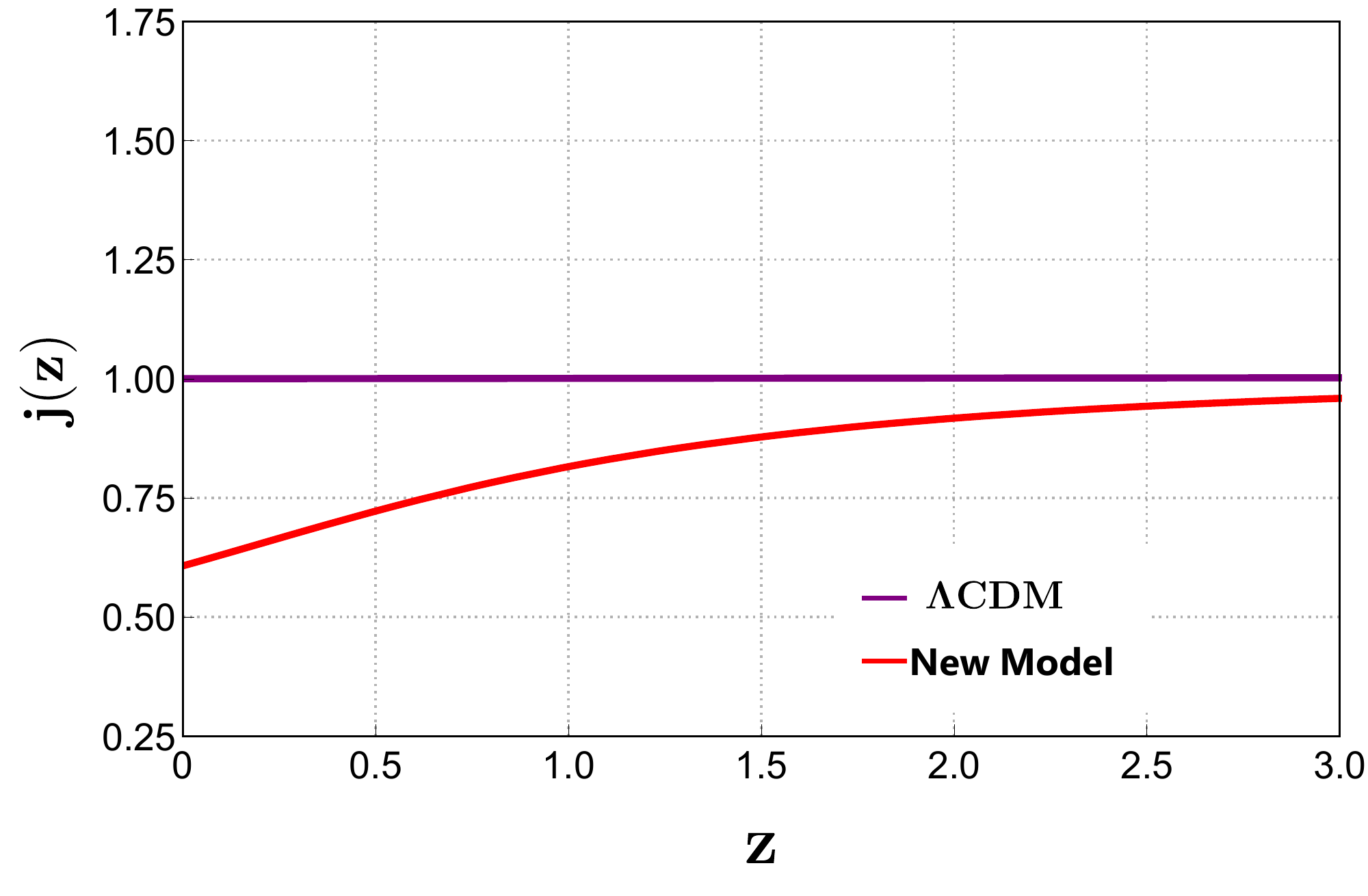}
\caption{Evolution of jerk parameter with of Model 4 with respect
to redshift.}\label{j(z) Model 4}
   \end{minipage}\hfill
   \begin{minipage}{0.49\textwidth}
     \centering
    \includegraphics[scale=0.42]{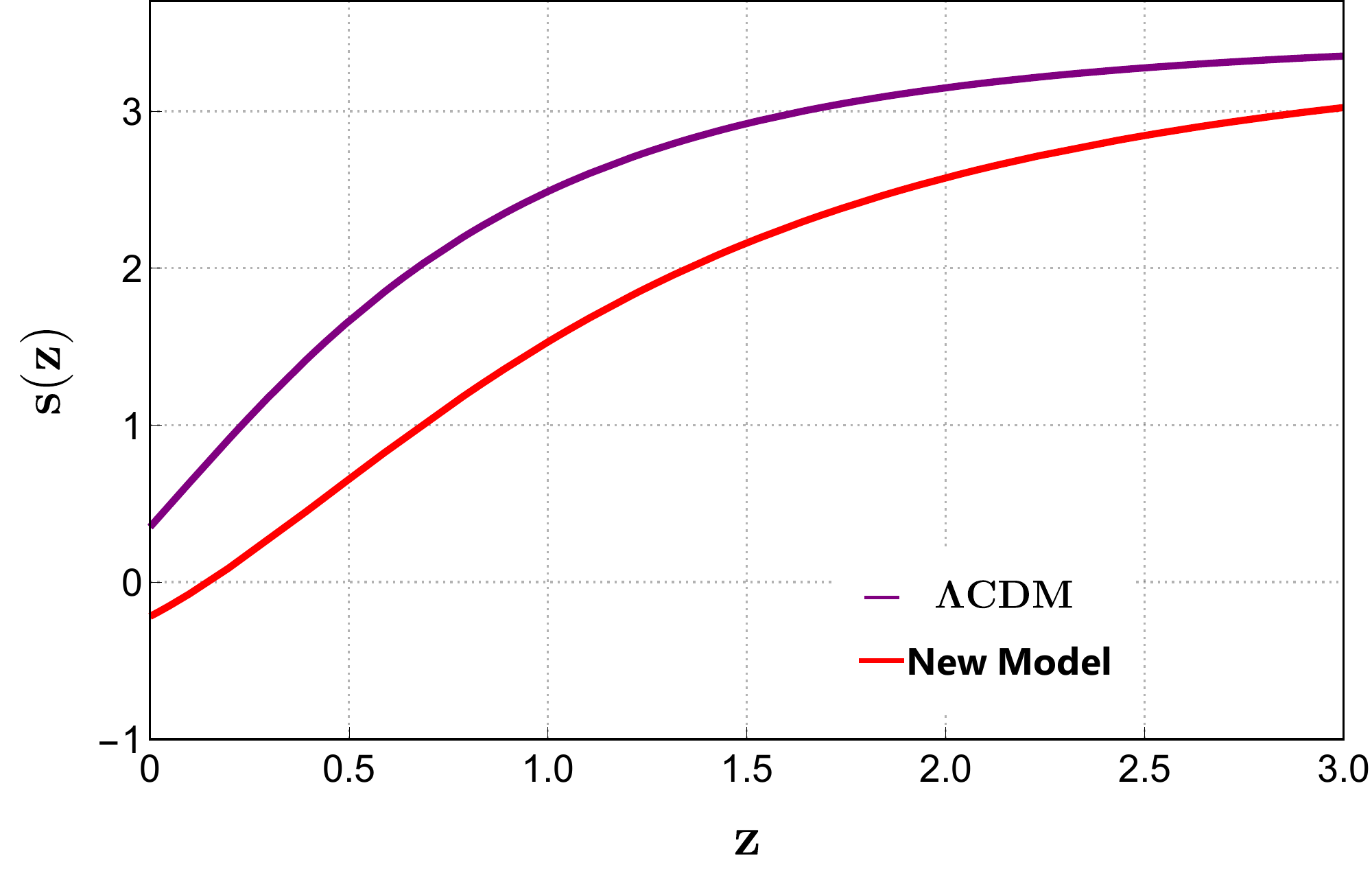}
\caption{Evolution of snap parameter with of Model 4 respect to
redshift.}\label{s(z) Model 4}
   \end{minipage}
\end{figure}

\newpage
\section{Detailed Description}\label{sec7}
\subsection{Cosmographic Analysis}

The cosmographic analysis provides a universal and effective way
to compare the solutions of the theoretical models with
cosmological observations. From the observational data, we obtain
a set of cosmological parameters, which must be compared with the
predicted values of the same parameters obtained from a given
model. The result of the comparison allows us to conclude the
acceptability of the considered model. Thus, for a complete
comparison of all models with the observations and the
$\Lambda$CDM model, we will consider an extended set of parameters
constructed from the higher-order time derivatives of the scale
factor. More exactly, we will concentrate on the comparative
behaviour of the deceleration, jerk, and snap parameters of all
models and $\Lambda$CDM models.

\subsubsection{Deceleration parameter}
While analyzing Model 1's trajectory, the behavior of the model's
deceleration parameter is nearly comparable to the $\Lambda$CDM
model in the redshift range of q $\in$ [-0,2], but Model 1 endures
a super acceleration in the future ( Fig: \ref{q(z) Model 1}).
Model 2 appears to have the same behavior as $\Lambda$CDM in the q
$\in$ [-0.8,4], but Model 2 is slower since it achieves the value
$-0.83565$ as $z \rightarrow 0$. (Fig: \ref{q(z) Model 2}). In
Model 3, this parameter appears to behave similarly to the
$\Lambda$CDM model, in q $\in$ [-0.5,6], before experiencing a
super acceleration in the near future as "q = -1.25464" (Fig:
\ref{q(z) Model 2}). Model 4 behaves differently from the
$\Lambda$CDM. but it aquire the same value as  $\Lambda$CDM in
near future \ref{q(z) Model 4}).

\subsubsection{The jerk parameter}
The jerk parameter of Model 1 basically different from
$\Lambda$CDM at high as well as low redshift. However, important
Model 1 predicts the higher value of $j = 1.357595$ at $z = 0$
(Fig: \ref{s(z) Model 1}), Meanwhile, on the other hand, Model 2
also shows different behaviour at both high and low redshift, and
seems to coincide with $\Lambda$CDM at redshift value of $z = 2$
and $z = 0.342566$, and this model also predicts the higher value
of $j = 1.134545$ (Fig: \ref{s(z) Model 2}). Model 3 shows
different behavior than the $\Lambda$CDM as at high redshift $ z >
2 $, it having a higher value than the $\Lambda$CDM of $j =
1.2657$, but it cuts the trajectory of $\Lambda$CDM, twice at $z =
1.5 $ and $z = 0.3$, finally, at lower redshift, it attains the
$j$ value of 1.05417, which is a higher than $\Lambda$CDM (Fig:
\ref{s(z) Model 3}). Although the jerk of Model 4 is inconsistent
with $\Lambda$CDM, since it shows different evolution at both high
and low redshifts and predicts the lower value of $j = 0.608251$
(Fig: \ref{s(z) Model 4}).

\subsubsection{The snap parameter}
This parameter of Model 1 and Model 2 is significantly systematic
the difference with $\Lambda$CDM within the whole redshift range,
but this parameter of Model 1 monotonically decreases in the
redshift range of $z > 0.3$  and acquires a sudden increase in "s"
value and predicts the higher value of $j = 1.608251$ then
$\Lambda$CDM, but Model 2 predicts the same value of $j$ as
$\Lambda$CDM,(Fig: \ref{s(z) Model 1}, \ref{s(z) Model 2}). Model
3 trajectory shows a proper systematic difference with
$\Lambda$CDM and predicts a higher value of "s" as 0.734546
\ref{s(z) Model 3} Finally, the snap trajectory of Model 4 is
notably non-identical with the $\Lambda$CDM, in the given redshift
range and accommodate the ``s" value of $-0.222743$ as
$z\rightarrow 0$. which is lower than $\Lambda$CDM (Fig: \ref{s(z)
Model 4}).

\section{Information Criteria}\label{sec8}
To discuss the viable model analysis, we need to know the study of
information criteria (IC). The Akaike Information Criteria (AIC)
\cite{52} is merely used among all ICs. The AIC is an
asymptotically unbiased estimator of Kullback-Leibler information
as the AIC is an approximate minimization of the Kullback-Leibler
information. The Gaussian estimator for the AIC can be written as
\cite{53,54,55,56} $\text{AIC}=-2\ln ({\cal
L}_{max})+2\kappa+\frac{2\kappa(\kappa+1)}{N-\kappa-1}$ where
${\cal L}_{max}$ is the maximum likelihood function, $\kappa$ is
the number of parameters of the models, and $N$ is the number of
data points used in the data fit of the models. Since for the
models, $N\gg 1$, so for this assumption, the above expression
converts to the original AIC like $\text{AIC}=-2\ln ({\cal
L}_{max})+2\kappa$. If the set of models is given, the deviations
of the IC values are reduced to 
$\triangle\text{AIC}=\text{AIC}_{model}-\text{AIC}_{min}=\triangle\chi^{2}_{min}+2\triangle\kappa$
In the study of data analysis, the more favorable range of
$\triangle\text{AIC}$ is $(0,2)$. The low favorable range of
$\triangle\text{AIC}$ is $(4,7)$, while
$\triangle\text{AIC}>10$ provides less support model.\\

\begin{table}[H]
\begin{center}
\begin{tabular}{|c|c|c|c|c|}
\hline
Model & $\chi_{min}^{2}$ & $\chi_{red}^{2}$ & $AIC$ & $\Delta AIC$  \\
\hline $\Lambda$CDM Model & 1102.67 & 0.981 & 1106.67 & 0  \\
\hline  Model 1 & 1103.21 & 0.961 & 1109.69 & 0.54   \\
\hline Model 2 & 1103.05 & 0.963 & 1107.05 & 0.38
\\ \hline Model 3 & 1103.85 & 0.965 & 1107.85 & 1.18 \\  \hline  Model 4 &
1103.76 & 0.972 & 1109.76 & 3.09 \\ \hline
\end{tabular}
\caption{Summary of the    $ {{\chi}^2_{min}}$,
${{\chi}^2_{red}}$,  $AIC$  and $\Delta AIC$.} \label{table3}
\end{center}
\end{table}

\section{Discussions and Conclusions}\label{sec9}
We have assumed the FLRW model of the universe in the presence of
radiation, dark matter and dark energy. Instead of considering the
well-known parametrized dark energy equation of state, we have
considered the analogous form of parametrized deceleration
parameter for the dark energy component and found the Hubble
parameter in terms of redshift with other model parameters. Here
we have assumed Model 1 (Wetterich type), Model 2 (Barboza-Alcaniz
type) and Model 3 (CPL type), which contains two unknown
parameters. Also, we have introduced a new Model 4 for
parametrized deceleration parameters, which also contains two
unknown parameters. The model parameters have been constrained for
$H(z)$ datasets, SNIa datasets, and BAO datasets by MCMC
method. Using the best-fit parameters, we have shown the nature of
the deceleration parameter, jerk parameter, and snap parameter. The
viability of the models has been studied by the information
criteria. We have compared all the models as well as compared with
the $\Lambda$CDM model (which is the base model) to get which
model is more viable than others. From Table: \ref{table3}, we
observe that Models 1 - 4 are all viable models, but (i) Model 3
is more viable than Model 4, (ii) Model 1 is more viable than
Model 3 and  (iii) Model 2 is more viable
than Model 1 compared to the $\Lambda$CDM model. \cite{l:2023rdx}  \\\\

{\bf Acknowledgement:} TR is thankful to IIEST, Shibpur, India for providing Institute fellowship (SRF). G. Mustafa is very thankful to Prof. Gao Xianlong from the Department of Physics, Zhejiang Normal University, for his kind support and help during this research. Further, G. Mustafa acknowledges Grant No. ZC304022919 to support his Postdoctoral Fellowship at Zhejiang Normal University.\\\\

\bibliographystyle{elsarticle-num}
\bibliography{mybib}

\clearpage
\begin{widetext}
\begin{table*}[htbp]
\begin{tabular}{|c|c|c|c|c|c|}
\hline
  \bf BAO name          &\bf$\textrm{redshift z}$  &\bf Experiment &\bf Measurement &$\bf {Standard deviation}$ &Ref. \\
\hline \multirow{1}{*} {6dFGS} &$0.106$ &${r_s}/{D_V}$  &$0.336$
&$0.015$ &\cite{100}  \\ [0.2cm]

\hline \multirow{1}{*} {SDSS DR7} &$0.15$
&${D_V}(r_{s,fidd}/r_{s})$ &$664$ &$25.0$ &\cite{101}   \\ [0.2cm]

\hline \multirow{1}{*} {SDSS-DR7 + 2dFGRS} &$0.275$
&${r_{s}}/{D_V}$ &$0.1390$ &$0.0037$ &\cite{102}   \\ [0.2cm]

\hline \multirow{1}{*} {SDSS-DR11 LOWZ} &$0.32$
&${D_V}(r_{d,fidd}/r_{s})$ &$1264$ &$25$ &\cite{103}    \\ [0.2cm]

\hline \multirow{1}{*} {SDSS-III DR8} &$0.54$ &${D_A}/{r_{s}}$
&$9.212$    &$0.41$ &\cite{104}   \\ [0.2cm]

\hline \multirow{1}{*} {SDSSIII/ DR9} &$0.57$ &${D_V}/{r_{s}}$
&$13.67$    &$0.22$ &\cite{105}   \\ [0.2cm]

\hline \multirow{1}{*} {SDSS-IV DR14} &$0.72$
&${D_V}(r_{s,fidd}/r_{s})$ &$2353$    &$63$ &\cite{107}    \\
[0.2cm]

\hline \multirow{1}{*} {DES Year 1} &$0.81$ &${D_A}/{r_{s}}$
&$10.75$    &$0.43$ &\cite{108}    \\ [0.2cm]

\hline \multirow{3}{*} {DECals DR8}      &$0.874$
&${D_A}(r_{s,fidd}/r_{s})$ &$1680$    &$109$  &\cite{107}  \\
[0.2cm]
                               &$0.697$ &${D_V}(r_{s,fidd}/r_{s})$ &$2353$    &$63$ &    \\ [0.2cm]
\hline \multirow{1}{*} {eBoss DR16 BAO+RSD} &$1.48$
&${D_H}.{r_{s}}$ &$13.23$    &$0.47$ &\cite{109}    \\ [0.2cm]

\hline \multirow{1}{*} {SDSS-IV/DR14} &$1.52$
&${D_V}(r_{s,fidd}/r_{s})$ &$3843$    &$147.0$ &\cite{108}  \\
[0.2cm]

\hline \multirow{1}{*} {Boss Lya quasars DR9} &$2.3$ &$H.{r_{s}}$
&$34188$    &$1188$ &\cite{111}   \\ [0.2cm]

\hline \multirow{1}{*} {BOSS DR14 Lya in LyBeta} &$2.34$
&${D_H}.{r_{s}}$ &$8.86$    &$0.29$  &\cite{112}  \\ [0.2cm]

\hline \multirow{3}{*} {WiggleZ}      &$0.44$  &   &$0.0870$
&$0.0042$   &  \\ [0.3cm]
                               &$0.6$  &${r_{s}}/{D_V}$    &$0.0672$ &$0.0031$  &\cite{113}\\[0.2cm]

                               &$0.73$  &  &$0.0593$ &$0.0020$ &   \\[0.3cm]

\hline
\end{tabular}
\caption{Summary of the Baryon Acoustic Oscillations measurements used in this work.}\label{tab_BAO}
\end{table*}
\end{widetext}

\end{document}